\documentclass[useAMS,usenatbib,usegraphicx]{mn2e}

\def\deg{\hbox{$^\circ$~\/}}
\def\degn{\hbox{$^\circ$\/}}
\def\Msun{\hbox{$\rm M_{\odot}$}}

\title[\textit{XMM-Newton} observations of  the Galactic Centre Region
  - I]{{\it XMM-Newton} observations of  the Galactic Centre Region - I:
  The distribution  of low-luminosity X-ray  sources} \author[V. Heard
  and R. S.  Warwick]{V. Heard$^{1}$\thanks{E-mail: vh41@le.ac.uk} and
  R.  S. Warwick$^{1}$\\  $^{1}$Department of  Physics  and Astronomy,
  University of Leicester, University Road, Leicester, UK}

\begin{document}

\date{Accepted 2012 October 24. Received 2012 October 8; in original form 2012 August 24}

\pagerange{\pageref{firstpage}--\pageref{lastpage}} \pubyear{2012}

\maketitle

\label{firstpage}

\begin{abstract}

\noindent We exploit \textit{XMM-Newton} archival data in a study
of the extended X-ray emission emanating from the Galactic Centre (GC) region.
\textit{XMM-Newton} EPIC-pn and EPIC-MOS observations, with a total
exposure time approaching
$0.5$ and $1$  Ms respectively, were used to create mosaiced images of
a  100 pc $\times$ 100  pc region centred on Sgr A* in  four bands covering
the  2--10  keV  energy range.  We   have  also  constructed  a  set  of
narrow-band images corresponding to the neutral iron fluorescence line
(Fe \textsc{i} K$\alpha$) at 6.4 keV and the K-shell lines at 6.7 keV
and 6.9 keV  from helium-like (Fe \textsc{xxv} K$\alpha$) and hydrogenic
(Fe \textsc{xxvi} Ly$\alpha$) iron ions.
We use a combination of  spatial and spectral  information to decompose
the GC X-ray  emission  into three  distinct  components. These comprise:
firstly the emission from  hard X-ray emitting unresolved point sources;
secondly the reflected continuum and fluorescent line emission
from dense molecular material; and, thirdly, the soft diffuse
emission from thermal plasma in the temperature range,
$kT \approx$ 0.8--1.5 keV.

We show that the unresolved-source component accounts for the bulk of the
6.7-keV and 6.9-keV line emission and also makes a major contribution to both
the 6.4-keV line emission and the 7.2--10 keV continuum flux. We fit the
observed X-ray surface brightness distribution with an empirical 2-d model,
which we then compare with a prediction based on an NIR-derived 3-d mass
model for the old stellar population in the GC. The  X-ray surface
brightness falls-off more rapidly with  angular offset from Sgr  A*
than the mass-model prediction.
One interpretation is  that the 2--10 keV X-ray emissivity increases
from $\approx 5 \times 10^{27}$ erg s$^{-1}~\Msun^{-1}$
at $20\arcmin$ up to almost twice this value at $2\arcmin$. Alternatively, some refinement of the mass model may be required, although it is unclear whether this applies to the Nuclear Stellar Cluster, the Nuclear Stellar Disc, or a combination of both components.

The unresolved hard X-ray emitting source population,  on the basis
of  spectral  comparisons,  is   most  likely  dominated  by  magnetic
cataclysmic  variables, primarily intermediate polars.
We  use  the  X-ray   observations  to  set
constraints on the number density of such sources at angular
offsets between $2\arcmin-20\arcmin$ from Sgr A* (projected distances
at the GC of 4.6--46 pc).  Our analysis does not support the conjecture
that a  significant fraction of the hard X-ray emission from the GC
originates in very-hot ($\sim7.5$ keV) {\it diffuse} thermal plasma.

\end{abstract}


\section{Introduction}
\label{sec:intro}

The centre of our Galaxy represents an intriguing laboratory for
the study of high energy processes and interactions. Within 
the central 100 pc, the rise in the stellar mass density is
complemented by a concentration of gaseous, largely molecular
matter, in a region known as the Central Molecular Zone \citep{morris96}.
Recent star formation coupled with supernova explosions every
few thousand years serve to inject high-energy particles into
the mix, which power non-thermal emission and drive
outflows (\citealt{crocker11}; and references therein).
At its heart, our Galaxy hosts a currently-quiescent
supermassive black hole, Sgr A*, with a mass of
$4 \times 10^6$ M$_{\odot}$  (\citealt{schodel02}; \citealt{ghez08}).
Quite plausibly recent activity on Sgr A* may also have energised
the surrounding region (\citealt{sunyaev98}; \citealt{koyama96}; 
\citealt{murakami00}; \citealt{ponti10}; \citealt{capelli12}). 
Given this context, it is not surprising that, even when the
contribution of highly-luminous X-ray binaries and transients is
excluded, the central region of our Galaxy exhibits a rich variety
of both compact X-ray sources and diffuse X-ray emission 
({\it e.g.}, \citealt{muno04a}; \citealt{park04}; 
\citealt{wang06}; \citealt{muno08}, \citeyear{muno09}).

One focus of research over the last forty years has been the nature
of the unresolved hard X-ray emission, known as the Galactic Ridge X-ray Emission (GRXE), which can be traced along
the Galactic Plane, reaching a maximum surface brightness
at the GC (\citealt{worrall83}; \citealt{warwick85}; 
\citealt{koyama86}, \citeyear{koyama89}; \citealt{revnivtsev06}).
This unresolved X-ray  emission is known to exhibit  many K-shell emission
lines  from highly-ionised atoms,  with the  most prominent  of these
lines  being  He-like  (Fe  \textsc{xxv} K$\alpha$)  and  H-like  (Fe
\textsc{xxvi} Ly$\alpha$) iron at 6.7 and 6.9 keV respectively. The flux
ratio  of  Fe   \textsc{xxv}  K$\alpha$/Fe  \textsc{xxvi} Ly$\alpha$
suggests the origin is in optically-thin thermal  plasma 
emission at  a temperature $\sim7.5$ keV 
(e.g., \citealt{koyama86};   \citealt{koyama07a};
\citealt{yamauchi09}). If this emission is diffuse in nature,
its temperature is too high for it to  remain gravitationally
bound to the Galactic
Plane and  it would instead escape  with a velocity  exceeding 1000 km
s$^{-1}$ (\citealt{worrall83}; \citealt{koyama86}). The kinetic energy
required  to  sustain  the  plasma ($\sim10^{42}$  erg  s$^{-1}$),  is
equivalent to the occurrence of at least one supernova explosion every
thirty  years  (\citealt{kaneda97};  \citealt{yamasaki97}).  This  far
exceeds current estimates of the supernova   rate   within   the
region. Furthermore, the spectra  of supernova remnants (SNRs) exhibit
thermal plasma components with  temperatures typically in the range of
$kT\sim$~0.2--0.8 keV and seldom, if ever, with $kT>3$ keV. 
These difficulties call into question whether the emission can be
truly diffuse in nature.

Early \textit{Chandra}  deep-field observations  revealed a large
population of faint point sources  within   the   GC  region   
(\citealt{wang02};
\citealt{muno03}). More recently, \citet{revnivtsev09} have reported that
within a particular deep-field observation ($l_{\textsc{ii}}, 
b_{\textsc{ii}} = +0.08\degn, -1.42\degn$), more than $\sim80$ per cent 
of the continuum and line emission around 6.7 keV can be
resolved into point sources. A
high concentration of discrete  sources is perhaps to be expected,
given that the central 2$\degn$ $\times$ 0.8$\degn$ region of  the Milky Way
contains $\sim1$ per cent  of the Galactic mass
(\citealt{launhardt02}). In fact,   \citet{muno06}    compared   the   spatial
distribution of the X-ray emission from point sources with the stellar
mass  distribution   model  derived  from  the   \textit{NIR}  map  of
\citet{launhardt02} and, within the limits of the statistics,  found a
good correlation between  them. They conclude  that  the  majority  of 
point  sources  with $L_{\rm{X}} \le 10^{34}$ erg s$^{-1}$ 
are likely to be magnetic cataclysmic variables (CVs).
In such systems, an accretion shock is generated above the
white dwarf surface, which heats the accreting material to 
temperatures $kT > 15$ keV. The resulting highly-ionised, optically-thin plasma cools and, eventually, settles onto the white dwarf.
The resulting X-ray spectrum comprises a complex blend of contributions
generated at differing temperatures, densities and optical depths
(e.g., \citealt{cropper99}; \citealt{yuasa10}). As well as a hard continuum, 
the X-ray spectra of magnetic CVs are characterised by a complex
of iron-K lines including strong He-like, H-like and fluorescent
components, the latter arising from relatively cold iron in states
from Fe \textsc{i} to \textsc{xvii} \citep{hellier04}.
Support for the conjecture that magnetic CVs represent
an important GC source population also comes from
studies of the hard X-ray emitting,  low-luminosity X-ray source population
found locally in our Galaxy \citep{sazonov06}; and the distribution
of the GRXE across the central quadrants
of the Galactic Plane \citep{revnivtsev06}.

\citet{uchiyama11}  recently  investigated   the
Fe  \textsc{xxv} K$\alpha$  emission from both the   
Galactic  Ridge  and  GC  regions  using
\textit{Suzaku} data and compared  the spatial profiles of the line
emission  with  the  stellar  mass  distribution  models  considered by
\citet{muno06}. They concluded that  the distribution of the line
emission  along the Galactic  Ridge is  consistent with  an unresolved
point source  origin but argued that an  additional component is  required to
account for the excess emission seen within the GC region. In contrast,
based on a detailed spectral analysis of \textit{Suzaku} observations
in the Galactic bulge region, \citet{yuasa12} concluded that virtually
all of the unresolved X-ray emission could be explained in terms of 
the integrated emission of two point source components, magnetic
accreting white dwarfs, which give rise to the hard emission
extending from 5--50 keV, and stellar coronal
sources which might account for the softer ($kT$ 1.2--1.5 keV) thermal
emission evident in the bulge fields.

In this  paper, we report on a detailed investigation of the spatial and
spectral characteristics of the X-ray emission observed within
the central 100 parsecs of the Galaxy based on data from
\textit{XMM-Newton}. The emphasis of our study
is on constraining the  model for the unresolved, hard X-ray emitting
sources and setting limits on what fraction of the emission might
reasonably be described as diffuse in nature.  In the next section,
we give details of the \textit{XMM-Newton} observations selected  for
this analysis and the procedures followed in the data reduction. We also
describe the
techniques employed to construct mosaiced X-ray images covering a
range of broad and narrow energy bands
and to extract representative X-ray spectra.
In \S3 we outline how the spatial and spectral information
can be modelled
in terms of three distinct emission components. We then discuss in \S4
the contribution that unresolved low-luminosity sources make to the 
GC X-ray emission and whether, in light of our results, there is any
requirement for truly diffuse, very-hot thermal plasma emission in
the GC. 
Finally \S5 gives a brief summary of our conclusions. Throughout
this work, the distance  to  the  GC  is  assumed
to be  8  kpc  \citep{gillessen09}.

\section{Observations and Data Reduction}
\label{sec:datared}

The  \textit{XMM-Newton}  Science Archive  (XSA)  was  used to  select
observations, publicly  available as  of  2011 July,  with
pointing positions  within 0.7$\degn$ of ($l_{\textsc{ii}},  
b_{\textsc{ii}}$) = ($0\degn, 0\degn$).  A minimum effective exposure of  
1000 s in at least one of the  three EPIC instruments  (\citealt{struder01}; 
\citealt{turner01}) was set  as a requirement.  Similarly only
observations with at least one  of  the EPIC  cameras
operating  in  either full-frame  (FF)  or
extended full-frame (EFF) mode  were considered. The majority of the
observations were carried out using the medium filter, although
either the thin or thick filter were deployed in some instances.  Table
\ref{tab:obs}  provides a  summary of  the set  of \textit{XMM-Newton}
observations used in the current analysis and related information.

\begin{table}
\begin{center}
\caption{Details of the  \textit{XMM-Newton} observations used in this
work. All observations were made  with a Medium filter in place unless
otherwise stated.}
\begin{tabular}{c c c c c}
\hline ObsID&Observation  date&
\multicolumn{3}{c}{Exposure time (s)}\\
&&MOS-1&MOS-2&pn\\   
\hline
0112970701&2000-09-11&23110&23116&16951\\
0112970401&2000-09-19&22871&22671&13712\\
0112970501&2000-09-21&14650&14758&4716\\
0112971001&2000-09-24&8500&8898&7356$^{\dag}$\\
0112971501&2001-04-01&6205&7242&2929\\
0112971901&2001-04-01&5215&5804&1339\\
0112972101&2001-09-04&22500&22515&16447\\
0111350101&2002-02-26&38944&39164&31139$^{\dag}$\\
0111350301&2002-10-03&8074&8097&5914$^{\dag}$\\
0202670501&2004-03-28&26554&29254&9064\\
0202670601&2004-03-30&34440&37048&24267\\
0202670701&2004-08-31&81716&84000&38114\\
0202670801&2004-09-02&102082&106098&62475\\
0302882501&2006-02-27&8167&8407&4595\\
0302882601&2006-02-27&2858&3153&966\\
0302882701&2006-02-27&4892&5097&-\\
0302882801&2006-02-27&7167&7072&2231\\
0302882901&2006-02-27&6667&6575&1876\\
0302883001&2006-02-27&6764&6268&-\\
0302883101&2006-02-27&10491&10101&1593\\
0302883201&2006-03-29&6132&6138&2541\\
0302883901&2006-09-08&6380&6285&2366\\
0302884001&2006-09-08&6377&6389&3319\\
0302884101&2006-09-08&6074&5683&1225\\
0302884201&2006-09-08&6480&6086&4363\\
0302884301&2006-09-09&6379&6181&4014\\
0302884401&2006-09-09&5440&5145&2748\\
0302884501&2006-09-09&8154&7763&5065\\
0506291201&2007-02-27&24438&19010&-\\
0402430701&2007-03-30&30390&29680&8827\\
0402430301&2007-04-01&52528&42381&16728\\
0402430401&2007-04-03&36809&31887&15841\\
0504940101&2007-09-06&6323&5918&2831\\
0504940201&2007-09-06&9696&9588&-\\
0504940401&2007-09-06&6117&6227&4396\\
0504940501&2007-09-06&5912&5929&4050\\
0504940601&2007-09-06&3552&3061&-\\
0504940701&2007-09-06&6419&5726&3355\\
0511000101&2008-03-03&1582$^{\ddag}$&1781$^{\ddag}$&-\\
0511000301&2008-03-03&5431$^{\ddag}$&5238$^{\ddag}$&606$^{\ddag}$\\
0511000501&2008-03-04&6020$^{\ddag}$&6229$^{\ddag}$&1528$^{\ddag}$\\
0511000701&2008-03-04&6422$^{\ddag}$&5828$^{\ddag}$&2825$^{\ddag}$\\
0511000901&2008-03-04&6028$^{\ddag}$&4828$^{\ddag}$&4303$^{\ddag}$\\
0511001101&2008-03-04&6518$^{\ddag}$&6529$^{\ddag}$&4352$^{\ddag}$\\
0511001301&2008-03-04&4730$^{\ddag}$&2950$^{\ddag}$&2342$^{\ddag}$\\
0505870301&2008-03-09&12275&12073&4704\\
0505670101&2008-03-23&79223&73233&44097\\
0511000201&2008-09-23&6319$^{\ddag}$&6325$^{\ddag}$&3856$^{\ddag}$\\
0511000401&2008-09-23&3688$^{\ddag}$&2481$^{\ddag}$&3682$^{\ddag}$\\
0511000601&2008-09-23&6518$^{\ddag}$&6328$^{\ddag}$&4198$^{\ddag}$\\
0511000801&2008-09-27&6409$^{\ddag}$&6224$^{\ddag}$&1931$^{\ddag}$\\
0511001001&2008-09-27&6518$^{\ddag}$&6529$^{\ddag}$&4352$^{\ddag}$\\
0511001201&2008-09-27&6518$^{\ddag}$&3522$^{\ddag}$&722$^{\ddag}$\\
0554750401&2009-04-01&28142&20968&24703\\
0554750501&2009-04-03&41733&39752&27885\\
0554750601&2009-04-05&36523&34838&23601\\     
\hline    
Total    time: &&942464&910071&487036\\
\hline
\end{tabular}
\label{tab:obs}
\end{center}
$^{\dag}$ Thick filter deployed.\\
$^{\ddag}$ Thin filter deployed.
\end{table}

The  data  reduction  was  based  on  \textsc{sas  v11.0}  (hereafter,
\textsc{sas}).  For  every observation listed  in Table \ref{tab:obs},
the Observation Data Files  for each of the EPIC instruments, pn,
MOS-1  and  MOS-2,  were  reprocessed  using  the  \textsc{sas}  tasks
\textsc{epchain} and  \textsc{emproc}.  Each dataset  was screened for
periods  of soft  proton flaring  through the  creation  of full-field
light curves  for the energy  range 10--12 keV (utilising  only single-pixel  events).  Good  Time  Interval (GTI)  selections  were made  by
setting a suitable threshold for each instrument, chosen to filter out
those  periods when the  instrument background  count rate  was highly
variable and clearly in excess of  the local quiescent level.   The
resulting
exposure times  for each instrument  after the application of  the GTI
filtering  are reported  in Table  \ref{tab:obs}.  The  total ``live''
time summed over the set of  50 pn observations and the 56 MOS-1/MOS-2
observations was, respectively, $\approx 0.5$ and $\approx 0.95$ Ms.
The  event lists  were  filtered so  as  to include  only single-  and
double-pixel   events   for  pn   (\texttt{PATTERN<=4})   and  up   to
quadruple-pixel  events for  MOS-1  and MOS-2  (\texttt{PATTERN<=12}).
Out-of-Time (OoT) events  have a non-negligible impact on  pn data and
appropriate  corrections were  applied.   As a  preliminary step,  the
\textsc{sas}  task \textsc{epchain}  was  used to  generate OoT  event
lists, filtered in an identical way to the primary datasets.

\subsection{Construction of the image mosaics}
\label{sec:images}

Images from each observation and from each camera were constructed 
in four ``broad'' bands  covering the 2--10 keV
bandpass (see  Table \ref{tab:bands}).  Instrument  exposure maps were
also made for each of these bands.  In  the case of the pn data, events with
energies in  the range 7.8--8.3  keV (which encompasses  the prominent
copper K$\alpha$  instrumental line) were  excluded when constructing
the  very-hard  band  image.   Similarly  for  the  MOS  very-hard
band, events in the 9.0--10.0  keV range were discarded (since the
MOS response rapidly declines in this energy range).

\begin{figure}
\centering
\includegraphics[width=70mm, angle=270]{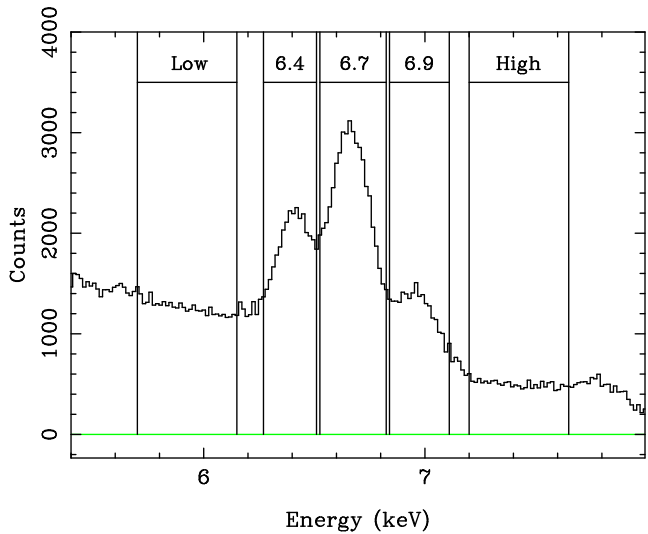}\\
\caption{The full-field spectrum from the pn camera (ObsID: 0202670801)
in the region of the Fe-K lines. The underlying continuum was sampled
in two narrow bands (marked ``Low'' and ``High'') at energies
immediately below and above the Fe-K complex. The various narrow bands
are delineated by the vertical lines.}
\label{fig:lines}
\end{figure}

Images  were  also made  in  a  number  of narrow  bands  encompassing
prominent Fe-K spectral  lines.  Table  \ref{tab:bands}  lists the  energy
ranges  used in  the present  study corresponding  to the  neutral (or
near-neutral) iron fluorescence line  (Fe \textsc{i} K$\alpha$) at 6.4
keV and  the K-shell  lines at 6.7 keV and  6.9 keV from  He-like (Fe
\textsc{xxv} K$\alpha$) and  hydrogenic (Fe \textsc{xxvi} Ly$\alpha$)
ions of iron. Two  narrow bands sampling the ``continuum'' at energies
immediately below and above the iron-line complex were also utilised (see Fig. 
\ref{fig:lines}).

\begin{table}
\begin{center}
\caption{Energy ranges (in keV) of the four broad bands covering the full
$2-10$  keV bandpass and of the  narrow bands encompassing the three
spectral lines of iron.}  
\renewcommand{\arraystretch}{0.75} \small
\begin{tabular}{c l}
\hline    
\multicolumn{2}{c}{Broad bands}\\   
\hline   
Medium&$2.0-4.5$\\
Hard&$4.5-6.0$\\  
Iron&$6.0-7.2$\\  
Very-hard&$7.2-10.0^{\P}$\\  
\hline
\multicolumn{2}{c}{Narrow bands}\\  
\hline 
Continuum-low&  $5.70-6.15$\\
Fe \textsc{i}    K$\alpha$  & $6.270-6.510$\\  
Fe \textsc{xxv}  K$\alpha$ & $6.525-6.825$\\  
Fe \textsc{xxvi} Ly$\alpha$ & $6.840-7.110$\\  
Continuum-high& $7.20-7.65$\\
\hline
\end{tabular}
\label{tab:bands}
\end{center}
$^{\P}$The $7.8-8.3$ keV spectral region encompassing the Cu K$\alpha$
instrumental line was excluded in the case of the pn data, whereas for
the MOS data the very-hard band was truncated at 9.0 keV.
\end{table}

The  first   correction  applied  to   the  pn  images   involved  the
construction of a set of OoT  images (one per band) from the OoT event
lists.   These OoT  images  were  scaled to  the  estimated OoT  event
fraction (6.3 and  2.3 per cent in FF and  EFF modes respectively) and
subtracted  from  the raw  pn  image  using  the \textsc{ftools}  task
\textsc{farith} (\textsc{ftools v6.11}, \citealt{blackburn95}).

The interactions  of high-energy  cosmic-ray particles, with  the EPIC
detectors  are  known  to  give  rise to  an  instrumental  background
consisting  of both a hard  continuum  and  fluorescent  line   features
\citep{lumb02}.  A  correction  for  the instrumental  background  was
applied to  both the pn  and MOS images  by making use of  EPIC filter
wheel closed (FWC) data  obtained from the background analysis website
maintained by ESA\footnote{
\texttt{http://xmm2.esac.esa.int/external/xmm$\_$sw$\_$cal/background/}\\\texttt{filter$\_$closed/index.shtml}.}.
With the  sky signal blocked,  the FWC exposures provide  a relatively
clean measure  of the instrumental  background and, therefore,  may be
used to  model and subtract  the instrument background.   Merged event
files containing all available FWC  data were selected for both the pn
FF  (Rev.   266--Rev.   2027)  and  pn  EFF  (Rev.   355--Rev.   1905)
modes. Merged FWC event files for the MOS cameras in FF mode were also
obtained with the  caveat that the FWC data for  MOS-1 were split into
two  categories:  before  (Rev.    231--Rev.   961)  and  after  (Rev.
961--Rev.  2027) the  failure of CCD-6 on 2005 March 9. The MOS-2
FWC event list  contained data taken from Rev.  230--Rev.  2027. Using
the  \textsc{skycast} software,  the FWC  data (initially  in detector
coordinates)  were  cast  into  images  in  the  sky-coordinate  frame
applicable to the particular observation.  The number of counts measured in
the corners of  the detector (\textit{i.e.}, the  regions of the detector
shielded from the sky) in  the actual observation were then divided by
the equivalent number of counts in the  FWC image to give a FWC scale factor
for  each band,  instrument,  and observation.   A  correction for  the
instrument  background  was  then   applied  to  each  observation  by
subtracting the scaled FWC image.

As a  final step prior to the  combining of the images  into a mosaic,
the images  were inspected visually for undesirable  artefacts such as
scattered light features and bright transient sources. These artefacts
were   spatially  masked  when   encountered and   a  corresponding
correction applied to the exposure maps.

In  the  production  of  the  image  mosaics,  observation  0202670801
(centred at RA, Dec of 266.4413$\degn$, -29.028972$\degn$) was used
as a  reference field. In  the case of  the pn camera, the  image data
deriving from  all the other observations were  resampled and co-added
to the reference  image.  This process was repeated  for each band and
also for each exposure map.  Flat-fielding then involved dividing each
co-added  image by  the appropriate  co-added exposure  map.   For the
narrow-band images  we used the  co-added 6.0--7.2 keV exposure  map in
this  flat-fielding step.   The  same  process was  used  for the  MOS
datasets, although  in this  case the MOS-1  and MOS-2 image  data and
exposure  maps were  combined prior  to the  flat-fielding  step.  The
resulting mosaiced  images have  a format of  $648 \times 648$ pixels
($4\arcsec$ pixels)    corresponding     to    a    $43.2\arcmin \times
43.2\arcmin$ field.  At the distance of the GC this field size covers a
projected extent of $100 \times 100$ pc.

\begin{figure*}
\centering
\begin{tabular}{cc}
\includegraphics[width=65mm, angle=270]{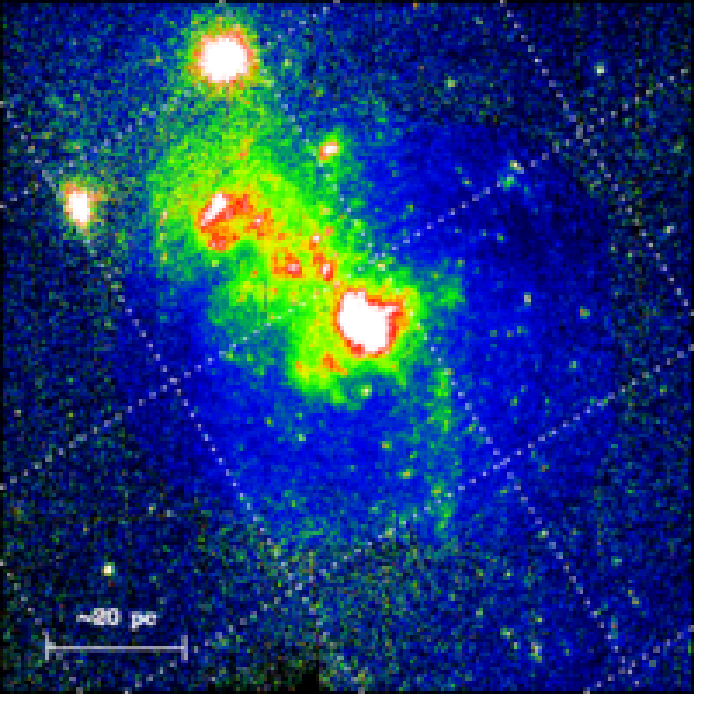}&
\includegraphics[width=65mm, angle=270]{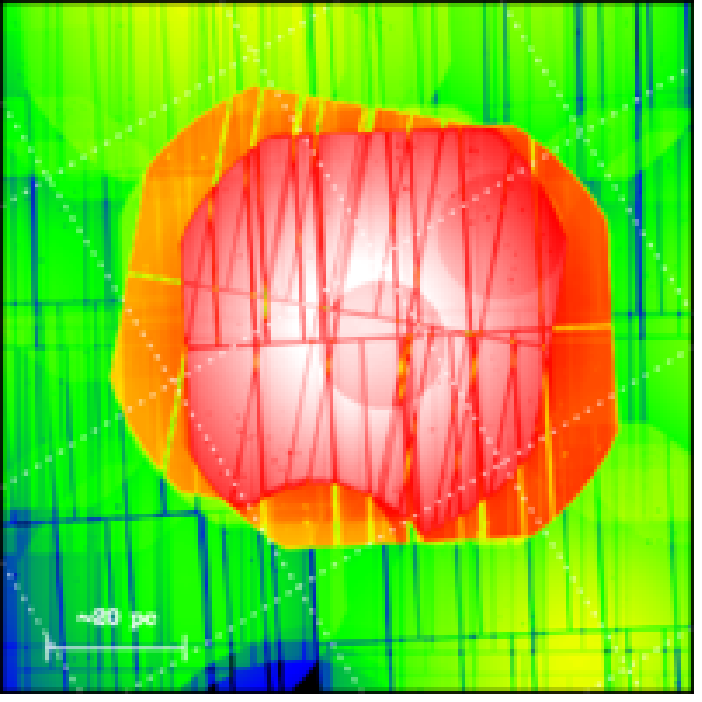}\\
\includegraphics[width=65mm, angle=0]{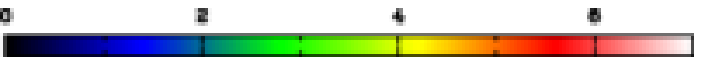}&
\includegraphics[width=65mm, angle=0]{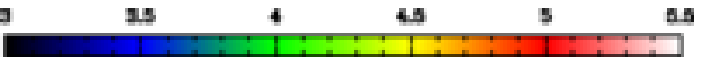}\\
\includegraphics[width=65mm, angle=270]{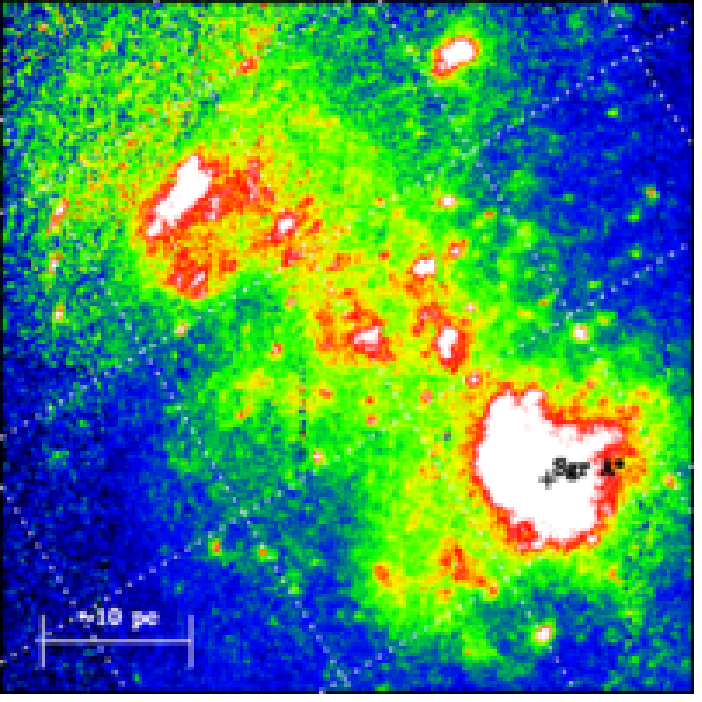}&
\includegraphics[width=65mm, angle=270]{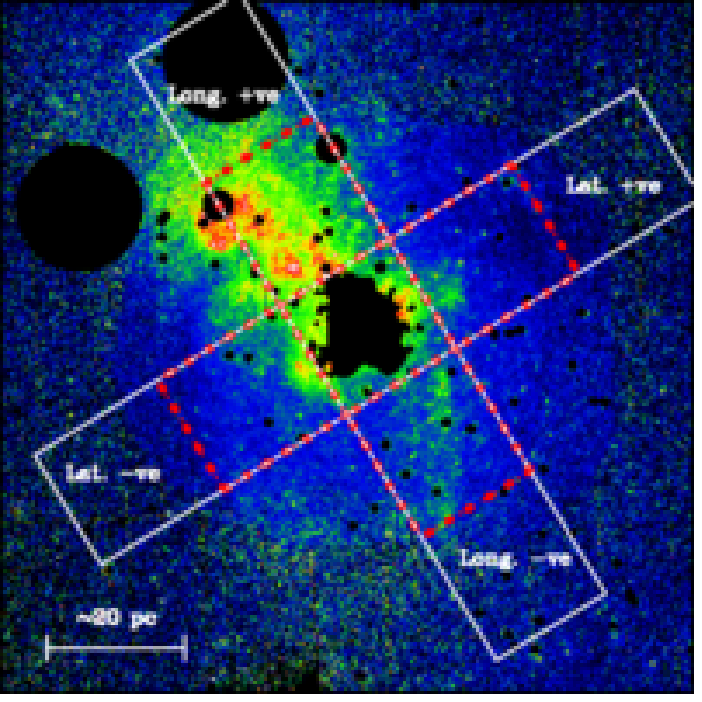}\\
\end{tabular}
\caption{Image  mosaics   based  on  \textit{XMM-Newton}  observations
covering   a  central   43.2\arcmin~$\times$   43.2\arcmin~field  (100
$\times$ 100  pc at the  GC). The images are constructed on a celestial
coordinate grid with east to the left.  
{\it Upper-left panel:} 2--10
keV  pn image. The adjacent colour bar is calibrated
in units of pn counts per pixel per 20 ks. {\it Upper-right panel:}  
The  corresponding pn exposure map. 
The adjacent colour bar is calibrated in units
of log$_{10}$ of the exposure (sec).  
{\it Lower-left panel:} A blow-up of the 2--10 keV image showing
the region to the north-east of Sgr A*.
{\it Lower-right panel:} The 2--10 keV pn
image overlaid by a  spatial mask which removes the most luminous
sources  (see   $\S$\ref{sec:masking}).   The  longitudinal  and
latitudinal  spatial cuts are represented
by the two intersecting white rectangles, which have long-axis dimensions
of $\pm 22\arcmin$ and full widths of $8\arcmin$ (see $\S$\ref{sec:cuts}).
The red boxes (of 
dimension $8\arcmin \times 9\arcmin$)  define the four
regions  from which spectral data were extracted (see $\S$\ref{sec:spec}).
In three of the panels the white dotted lines represent a Galactic
coordinate grid with a $15\arcmin$ spacing.}
\label{fig:ims}
\end{figure*}

The upper-left panel of  Fig. \ref{fig:ims} shows  the 2.0--10.0
keV  pn  mosaic  produced  by  summing the four broad-band  images.   The
corresponding   exposure  map   is   shown  in   the upper-right
panel. Fig. \ref{fig:ims} (lower-left panel) also presents a blow-up
of the  region to  the north-east of  Sgr A*,  where there is  extended high
surface brightness X-ray emission.

\begin{table}
\begin{center}
\caption{Factors  ($f$)  used   in  interpolating  between the continuum-low
and continuum-high  narrow band
measurements  to  give  the  continuum  underlying  the  line  image.}
\renewcommand{\arraystretch}{0.75} \small
\begin{tabular}{c c c}
\hline 
Line &pn&MOS\\
\hline
Fe 6.4-keV &0.72&0.64\\
Fe 6.7-keV &0.53&0.435\\
Fe 6.9-keV &0.325&0.245\\ 
\hline
\end{tabular}
\label{tab:factors}
\end{center}
\end{table}

\begin{table}
\begin{center}
\caption{Percentage spillover  from one iron-line band to another  as a
result  of  the limited  spectral  resolution  of  the EPIC  cameras.}
\renewcommand{\arraystretch}{0.75} \small
\begin{tabular}{c c c}
\hline &pn&MOS\\ \hline Fe 6.4-keV to Fe 6.7-keV &8.5&2.7\\ Fe 6.4-keV
to Fe 6.9-keV &5.0&4.1\\ Fe 6.7-keV to Fe 6.4-keV &2.2&1.9\\ \hline
\end{tabular}
\label{tab:kakbcont}
\end{center}
\end{table}

In  order to trace  the distribution  of the iron-line emission,  it is
necessary to make a correction  to the narrow-band images for the
in-band continuum  emission underlying  the emission lines.   Here the
approach  was  to calculate  an  interpolated  ``continuum'' image  by
multiplying the  narrow-band continuum-low and -high  images (see Table
\ref{tab:bands})  by a  factor $f$  and $1-f$  respectively,  and then
summing the  two products. Full-field spectra extracted  from a number
of the GC observations (see Fig. \ref{fig:lines}) were used to estimate
appropriate values of $f$ for the  three Fe-lines, 
for both  the pn and MOS  channels (see Table
\ref{tab:factors}).   Continuum-subtracted line
images were produced by subtracting the resultant interpolated
continuum images from the line images (with the former scaled so as to
account for the different bandpasses of the narrow-band
line and continuum images.

The need for a further correction to the line images arises due to 
the limited
spectral resolution of  the pn and MOS cameras.   This results in some
spillover of signal  from a given iron line into  the adjacent line band
or  bands (see Fig. \ref{fig:lines}).    
Here  we  used   the  X-ray  spectral-fitting  package
(\textsc{xspec}, \textsc{v12.6.0})  to estimate the spillover factors
for spectra  dominated either  by reflection or  hot-thermal emission.
The former may be modelled as a bright Fe \textsc{i} K$\alpha$ line at
6.4 keV  plus an associated Fe \textsc{i} K$\beta$ line  at 7.11 keV
(with 11 per  cent of the K$\alpha$ normalisation)  superimposed on a
power-law  continuum with  spectral slope  $\Gamma \approx  1.7$. Our
simulations  showed  that the  two  reflection  lines contaminate  the
6.7-keV and 6.9-keV narrow-bands as  summarised in the first two lines
of Table \ref{tab:kakbcont} (separate values are quoted for the
pn  and MOS instruments since their spectral resolution and
energy response differ).     Similarly,    the
spillover of the 6.7-keV He-like Fe K$\alpha$ emission  into the 6.4-keV
narrow-band  was simulated using  was a  solar-abundance \texttt{apec}
model   with   the   plasma    temperature   set   at   $kT=7.5$   keV
(\citealt{smith01}); the  results are  given in
the third line in Table \ref{tab:kakbcont}.  Final Fe-line images were
produced  by  subtracting  appropriate  fractions of  the  other  line
images.  Hereafter, we will often refer to the Fe \textsc{i} K$\alpha$, 
Fe \textsc{xxv} K$\alpha$ and Fe \textsc{xxvi}  Ly$\alpha$ lines
simply as the 6.4-keV, 6.7-keV and 6.9-keV lines, respectively.

\begin{figure*}
\centering
\begin{tabular}{cc}
\includegraphics[width=65mm,angle=0]{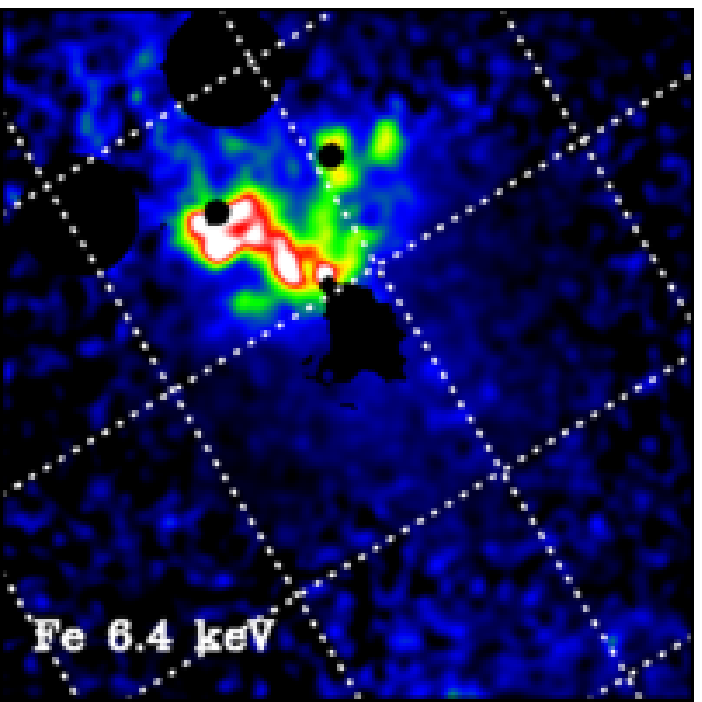}&
\includegraphics[width=65mm,angle=0]{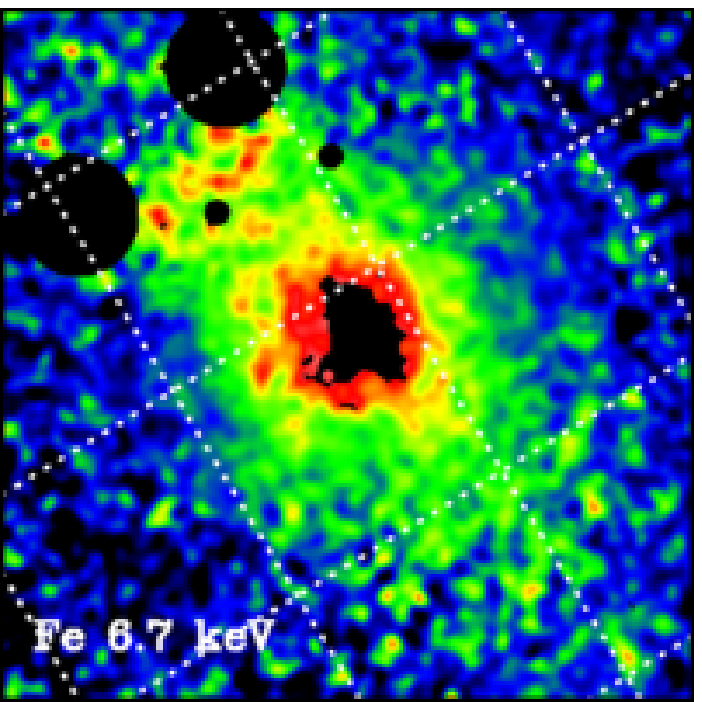}\\
\includegraphics[width=65mm,angle=0]{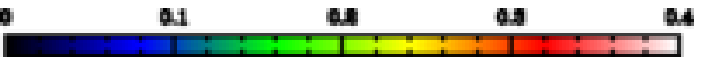}&
\includegraphics[width=65mm,angle=0]{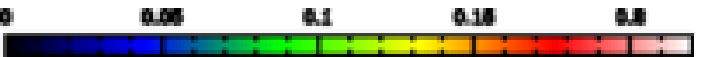}\\
\includegraphics[width=65mm,angle=0]{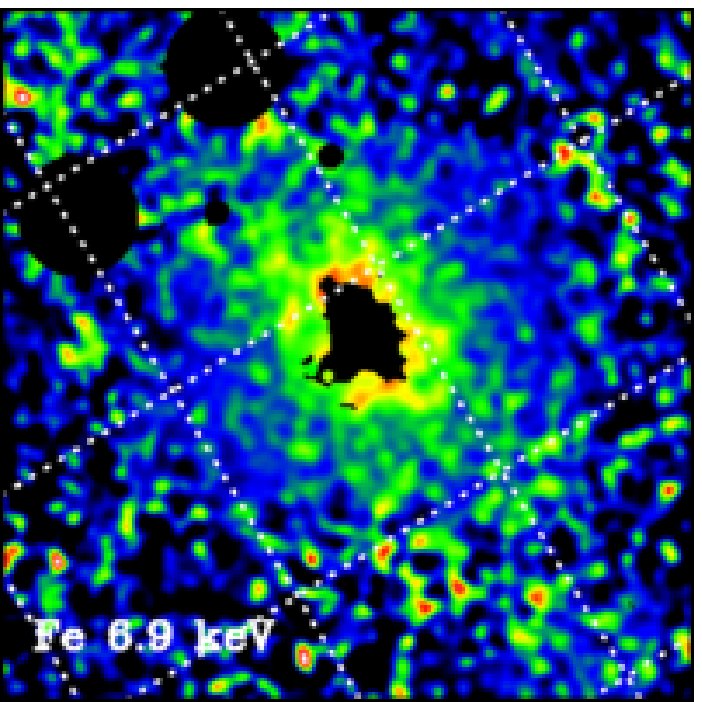}&
\includegraphics[width=65mm,angle=0]{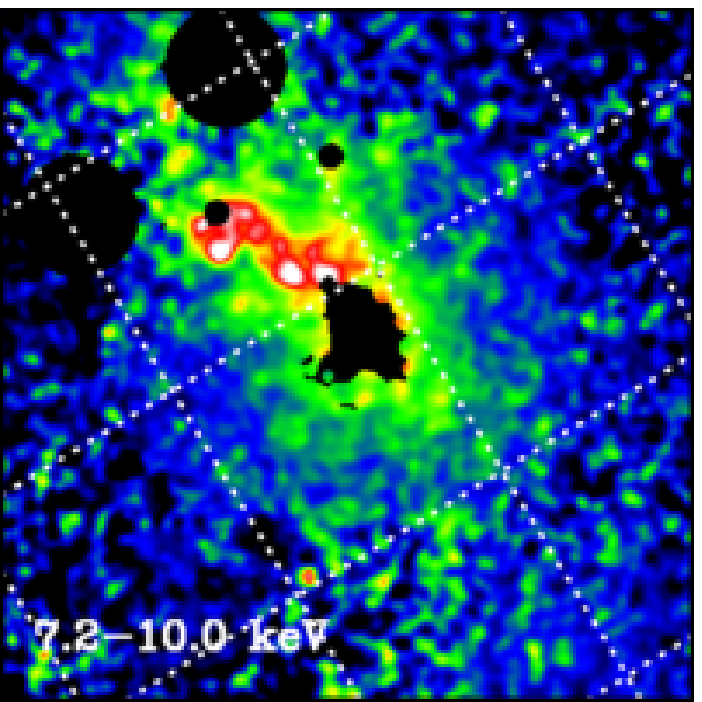}\\
\includegraphics[width=65mm,angle=0]{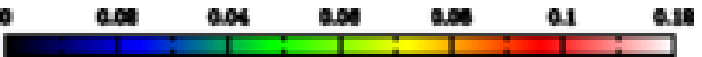}&
\includegraphics[width=65mm,angle=0]{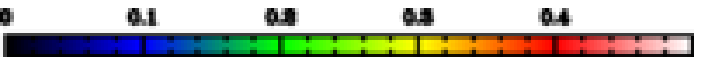}\\
\end{tabular}
\caption{The distribution of iron-line emission and the associated
very-hard continuum in the  central  100 pc  of the  Milky
Way.  {\it Upper-left panel:} the neutral iron fluorescence line
at 6.4 keV.
{\it Upper-right panel:} the He-like iron line (Fe \textsc{xxv} K$\alpha$)
at 6.7 keV. {\it Lower-left panel:} the H-like iron line (Fe \textsc{xxvi} Ly$\alpha$)
at 6.9 keV.
{\it Lower-right panel:} the very-hard (7.2--10 keV) band image.  
All four images have been smoothed with a Gaussian mask function of width 
$\sigma=5$ pixels. The adjacent colour bars are calibrated in units of pn counts per pixel per 20 ks. The white dotted lines represent a Galactic
coordinate grid with a $15\arcmin$ spacing.
}
\label{fig:lineims}
\end{figure*}

Fig. \ref{fig:lineims} shows the pn continuum-subtracted images for
the three iron lines. The  lower-right panel of Fig. \ref{fig:lineims}
also shows the very-hard (7.2--10.0 keV) band image.  
Note that all of these images have been heavily smoothed with
a circular Gaussian function of width $\sigma=5$ pixels.
The brightest 6.4-keV line emission is  concentrated to the north-east
of Sgr A* in a region with a high molecular cloud density
(\citealt{koyama96}; \citealt{park04}; \citealt{ponti10}; \citealt{capelli11}).
In contrast, the surface
brightness in the  6.7-keV  and 6.9-keV  iron-line images show a
smooth increase towards Sgr A*, as might be expected if these
lines arise predominantly in the integrated emission of
unresolved point sources
(\citealp{muno06}, \citeyear{muno09}; \citealt{revnivtsev07}; 
\citealt{hong09}). On the other hand, the very-hard band image
can be readily interpreted in terms of a combination of the two emission
distributions noted previously.

\subsection{Spatial masking of point sources}
\label{sec:masking}
The presence of luminous point sources within the field of view introduces
a degree of confusion noise into the images, which can be removed  via
spatial masking. 
In  order to produce  such  a  spatial   mask,  a  catalogue  of sources
was constructed based on  the Second \textit{XMM-Newton}  Serendipitous  Source
Catalogue   (the 2XMMi catalogue; \citealt{watson09}). Specifically, 
2XMMi sources within the GC field with 2.0--12.0 keV 
flux greater than $\sim1.5 \times 10^{-13}$ erg s$^{-1}$ cm$^{-2}$ 
were  included in a preliminary  source list. The mosaiced images were
then inspected at the source locations and a number of possibly spurious
sources discarded.
Conversely, a small number of sources  not included in  the 2XMMi  catalogue, 
but clearly visible  in the  images, were added  manually to the  list.
The result was a catalogue of 84 sources. A spatial mask was then
made by excising circular regions of radius $25\arcsec$ at each source
position (corresponding to the removal of roughly 80 per cent of the source
response). Larger excision circles were used for
two very bright sources to the north-east of Sgr A*, for the Arches cluster
and for the pulsar wind nebula, G0.13-0.11. We also excluded the
brightest regions of the Sgr A East complex (using a contour based
on its 6.7-keV Fe-line emission). The full mask array was then applied to
every image employed in our study prior to any smoothing;
the lower-right panel of Fig. \ref{fig:ims} shows the result
in the case of the 2--10 keV pn image.

\subsection{Spatial cuts}
\label{sec:cuts}

Our investigation of the various components of the diffuse X-ray emission
present in the GC requires both a spatial and spectral modelling approach.
We based our spatial modelling on cuts through the various images
in Galactic longitude and latitude centred on the position of Sgr A*.
The  white rectangles in Fig. \ref{fig:ims}
(lower-right panel) indicate the regions of the mosaiced images
which were thereby sampled.  The cuts extend out to offset angles
of $\pm22\arcmin$ with a $\pm4\arcmin$ extent perpendicular to the
cut axis. The spatial resolution along each cut is $1\arcmin$. 
We excluded segments of the cuts
for which more than half the area was already excluded by the source mask.
This resulted in the loss of the central $\pm2\arcmin$ region 
(which is confused by Sgr A East) in both cuts and the
eastern-most $4\arcmin$ section of the longitudinal cut.
The results from this analysis are presented in \S \ref{sec:cuts2}.

\subsection{Spectral extraction}
\label{sec:spec}

X-ray spectra were extracted from four regions within  the central 
100  pc zone, corresponding to the sky areas sampled by the spatial
cuts at offsets from Sgr A* of between $4\arcmin$ and $13\arcmin$  
(see the lower-right panel of Fig. \ref{fig:ims}).
The dimensions of each region are therefore $8\arcmin \times 9\arcmin$, 
with the longer axis orientated along the cut direction. Luminous point
sources were excluded from the extracted regions corresponding to those
masked out in $\S$\ref{sec:masking}.

For simplicity,  we extracted the spectra from a single  pn  observation  
(0202670801) for which the  exposure  time  after GTI-filtering 
was $\sim62$ ks.
The wide extent of the diffuse emission in the GC
field makes the selection of a background region for use in spectral analysis
problematic. We opted, therefore, 
to determine the background spectra from the appropriate FWC dataset. 
FWC spectra
were selected from geometrical areas on the detector matching those 
of the four sky regions. A simple scaling of the FWC spectrum  so as to match
the sky exposure then proved adequate as  the instrumental
background.
Subsequent spectral analysis then requires all the relevant GC emission
components as well as the  cosmic X-ray  background  (CXB) to be
included in the spectral model (see $\S$\ref{sec:spec2}). 
 The \textsc{sas} tools \textsc{arfgen} and \textsc{rmfgen}
were also used  to produce the appropriate auxiliary  response and response
matrix files. The spectra were grouped so that each contained at least
ten counts per bin.

\section{Spatial and spectral modelling of the GC emission}
\label{sec:props}

\subsection{Distinct emission components}
\label{sec:distinct}

A number of unresolved X-ray emitting components are evident in the GC region,
which can be distinguished by virtue of their distinctive spatial and spectral
characteristics. Here, for simplicity, we attempt to model
this emission in terms of just three major components. 

The first is the emission associated with the unresolved hard X-ray emitting
point sources. In the
present analysis we have masked out the sources brighter than
$1.5 \times10^{-13}$ erg s$^{-1}$ cm$^{-2}$ in the 2--12 keV band which,
at the distance of the GC, corresponds to a 2--10 keV X-ray luminosity 
threshold of 
$\approx 10^{33}$ erg s$^{-1}$. Hence here we are dealing with 
the integrated emission of sources less luminous than this.  
Our spectral description of this component is in terms of a hard
continuum (taken to be thermal bremsstrahlung
with $kT = 7.5$ keV) with associated Fe-line
emission at 6.4, 6.7 and 6.9 keV (\textit{i.e.} neutral, He-like and 
H-like iron K$\alpha$ emission).  We assume that unresolved sources give
rise to the bright central concentration of emission, which is most clearly
discerned in the 6.7-keV and 6.9-keV images (Fig. \ref{fig:lineims}).

The second major emission component observed in the GC is that associated
with dense molecular clouds. The most obvious signature is the
iron fluorescence line at 6.4 keV (\citealt{koyama07b}, \citeyear{koyama09}; \citealt{park04}; 
\citealt{ponti10}; \citealt{capelli12}). This fluorescence is
very likely excited by hard X-ray illumination of the clouds.
The Thomson scattering of the incident X-ray flux back into our
line-of-sight gives rise to a second spectral characteristic,
namely the presence of a hard power-law X-ray continuum (with 
photon index $\Gamma$ in the range 1.5--2.0).
Dense molecular material is known to be very unevenly distributed in the
GC  with a concentration to the north-east of Sgr A* ({\it e.g.,}
\citealt{tsuboi99}). This explains the highly asymmetric
surface brightness distribution seen in the 6.4-keV image, 
which is also reproduced in the very-hard band image 
(see Fig. \ref{fig:lineims}). 

A third emitting component is that arising from diffuse thermal plasma, 
most likely energised by recent supernova explosions and, in some localised
regions, by the colliding stellar winds of massive stars. This relatively soft
emission component is pervasive throughout the GC region; however, the
highest surface brightness region outside of the central few
arcmins, where emission associated with the Central Cluster and the Sgr A East SNR dominates (\citealt{baganoff03}; \citealt{muno04b}), is again located to the
north-east of Sgr A*, where the density of molecular material is high. 
The temperature and, possibly,  metal abundance of this soft thermal plasma
may vary with location. In modelling this component we assume that the thermal
structure can be adequately represented by two temperature components, one at
$kT \approx 0.8$ keV and the other at $kT \approx 1.5$ keV.  The former
accounts for much of the line emission evident in the X-ray spectrum of the
GC region below $\sim4$ keV, the most prominent spectral line being 
the K$\alpha$
line of He-like sulphur at 2.4 keV. The 1.5-keV component, whilst also
contributing significantly to the spectrum below 4 keV, is hot enough 
to produce 6.7-keV iron-line emission.

This paper focuses on an investigation of the spatial distribution 
and spectral characteristics of the emission associated with the
unresolved point source population.  An analysis of the softer thermal
plasma emission will follow in a later publication 
(Heard \& Warwick, {\it in preparation}). 

\begin{figure*}
\centering
\begin{tabular}{cc}
\includegraphics[width=45mm, angle=270]{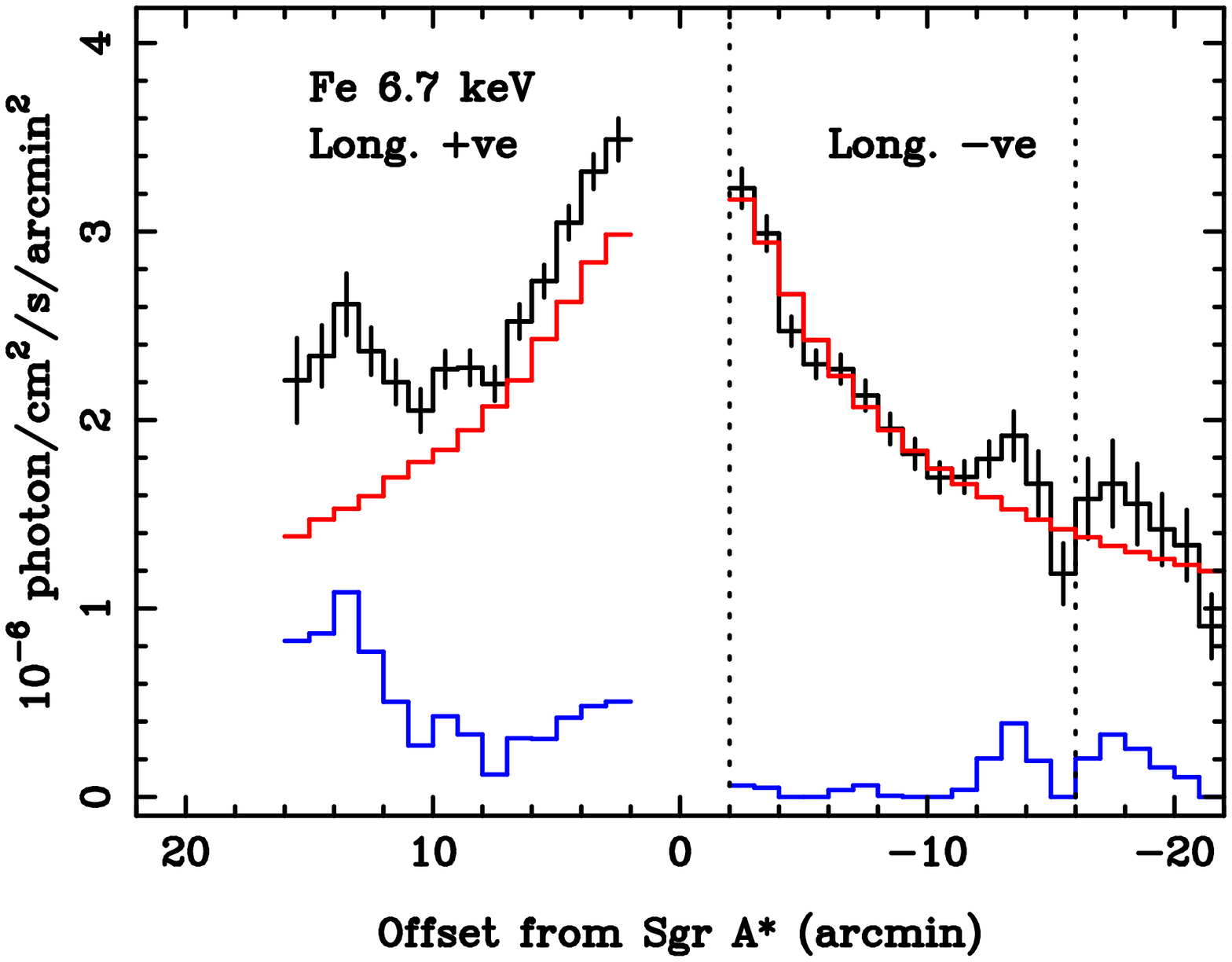}&
\includegraphics[width=45mm, angle=270]{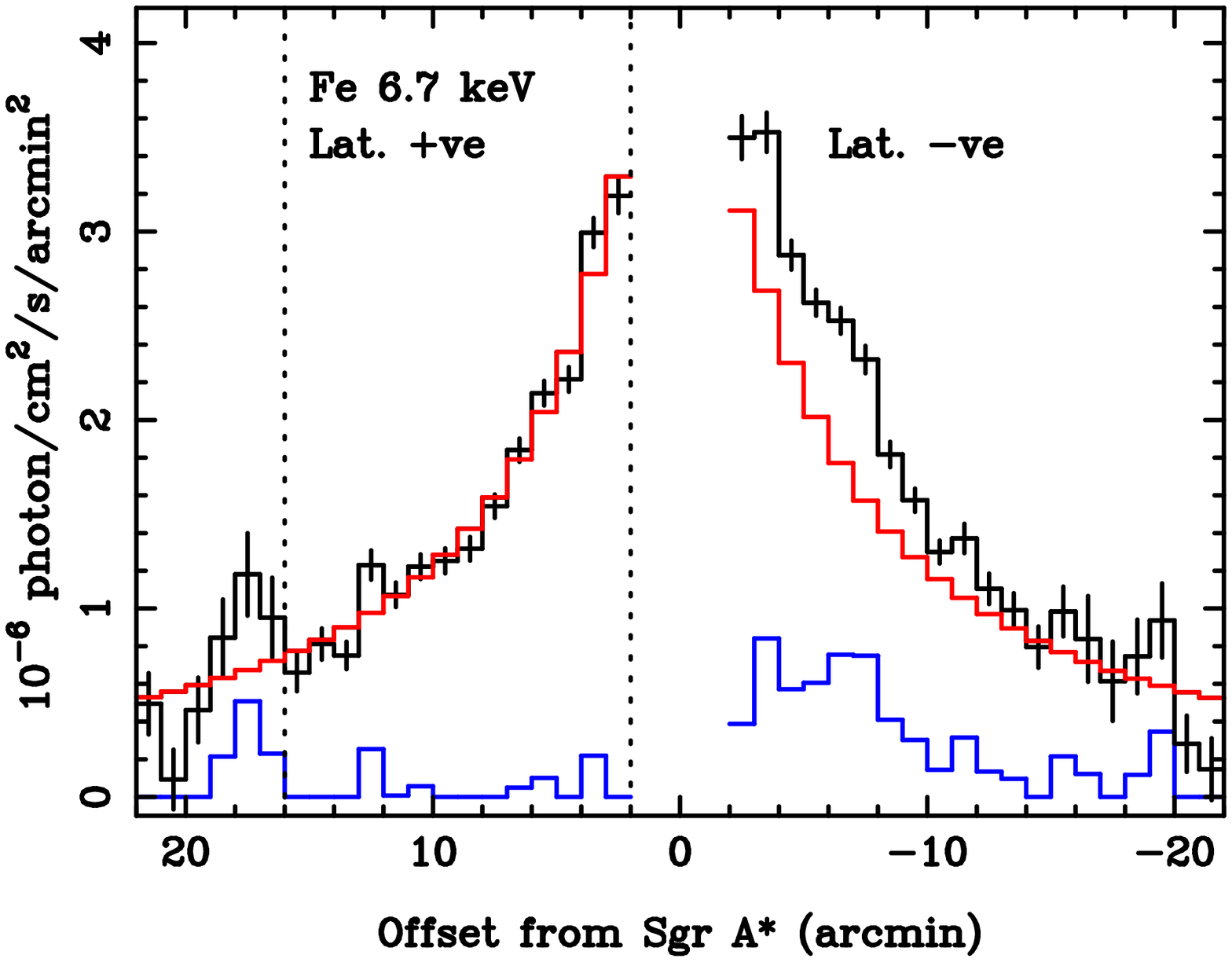}\\
\includegraphics[width=45mm, angle=270]{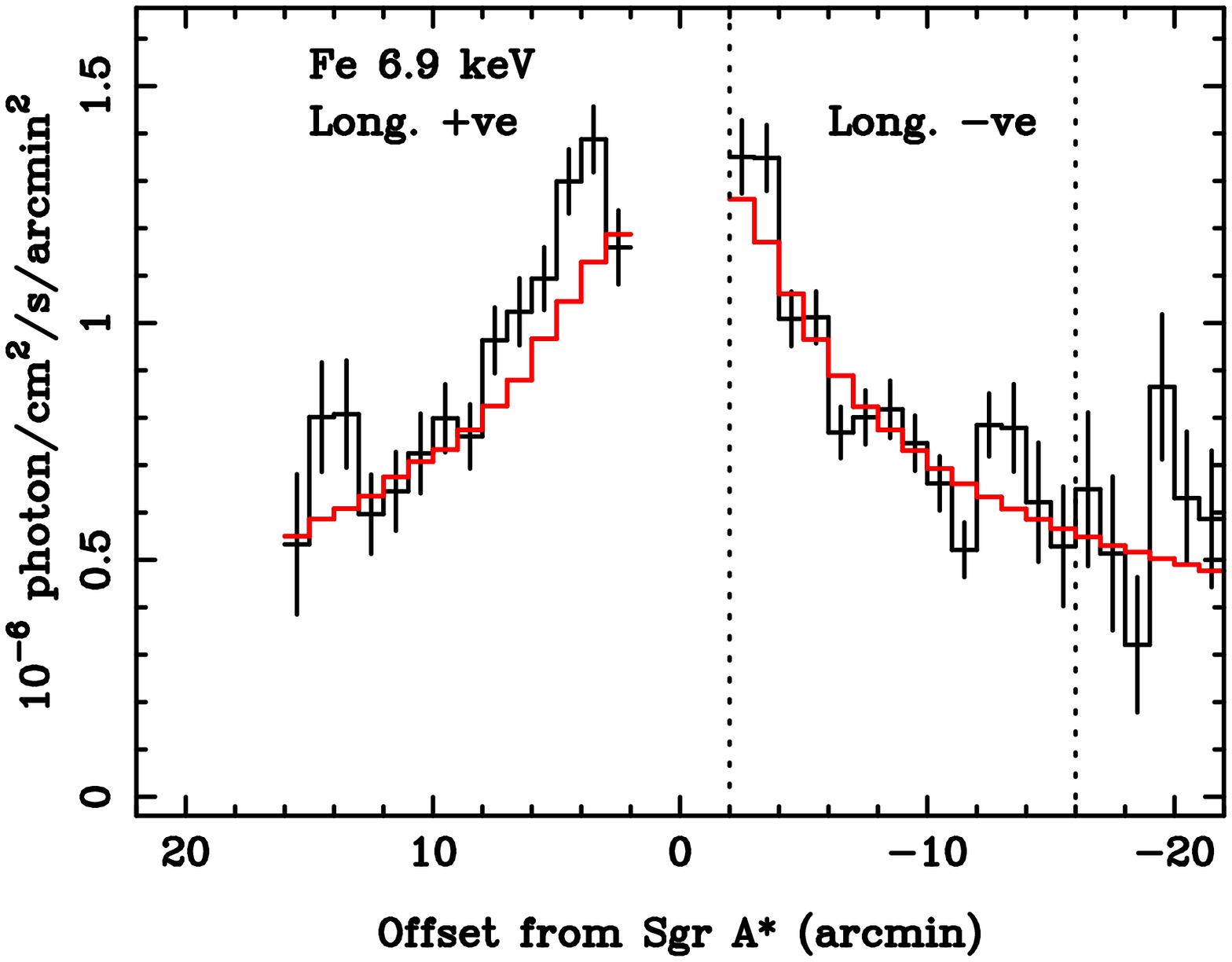}&
\includegraphics[width=45mm, angle=270]{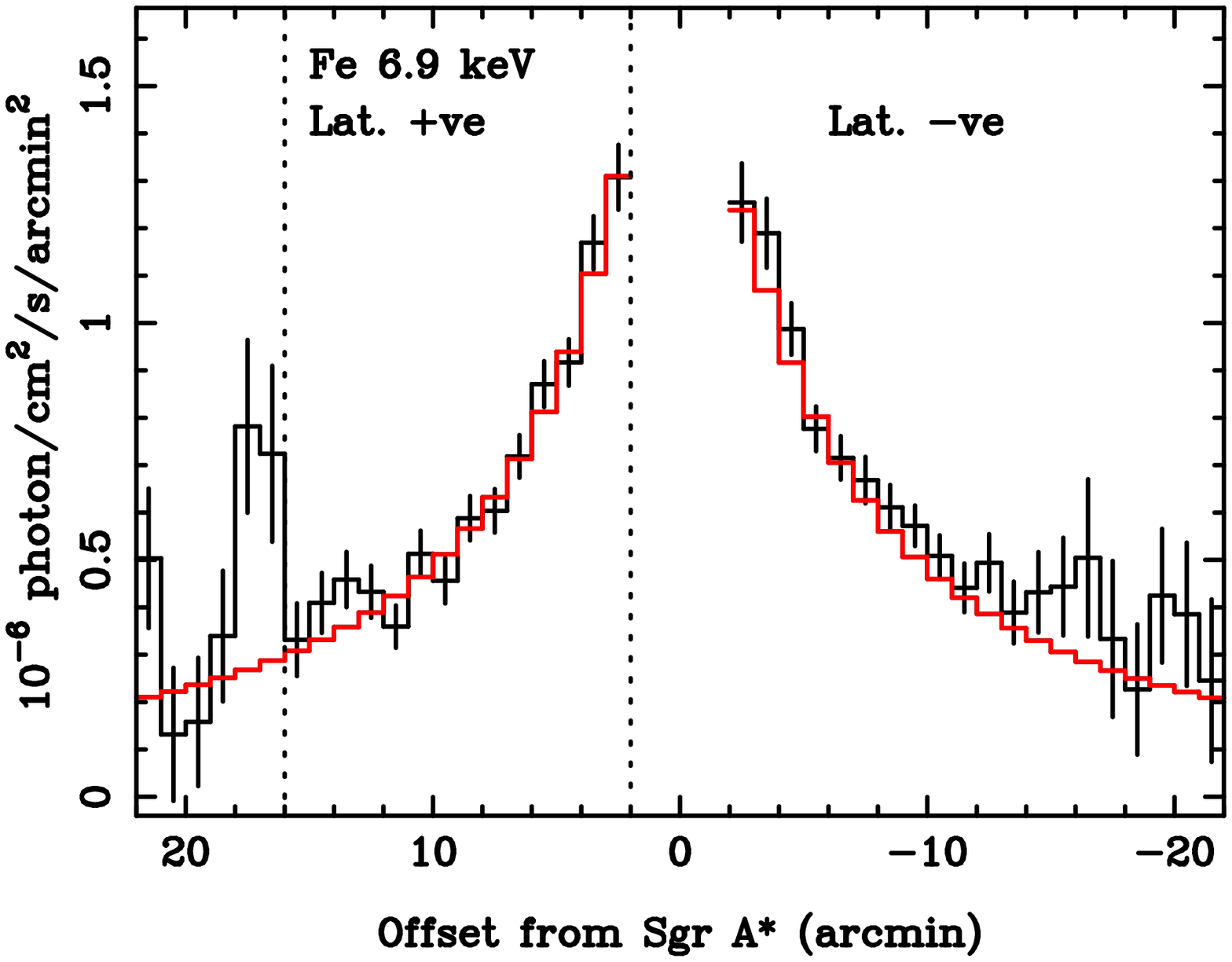} \\
\includegraphics[width=45mm, angle=270]{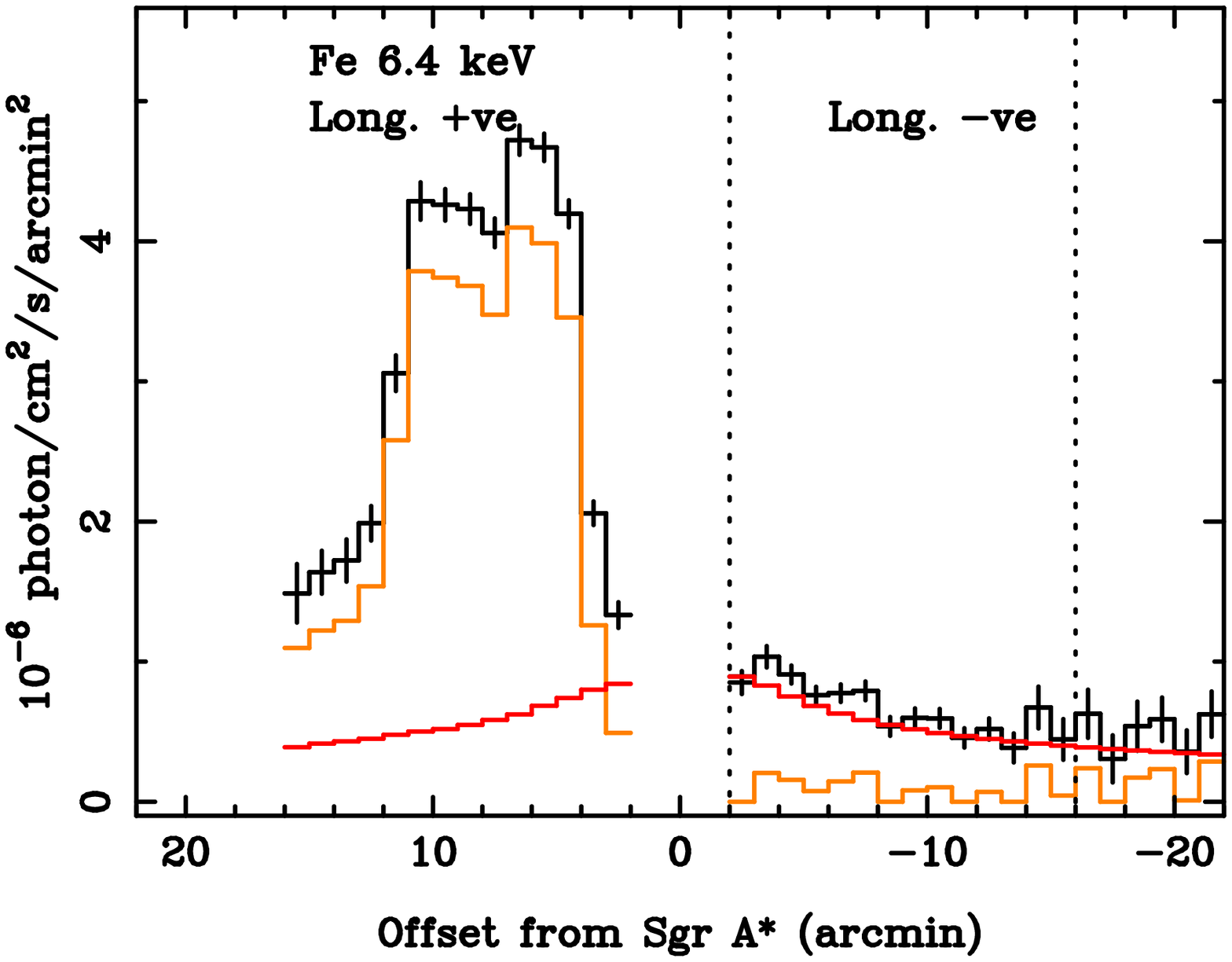}&
\includegraphics[width=45mm, angle=270]{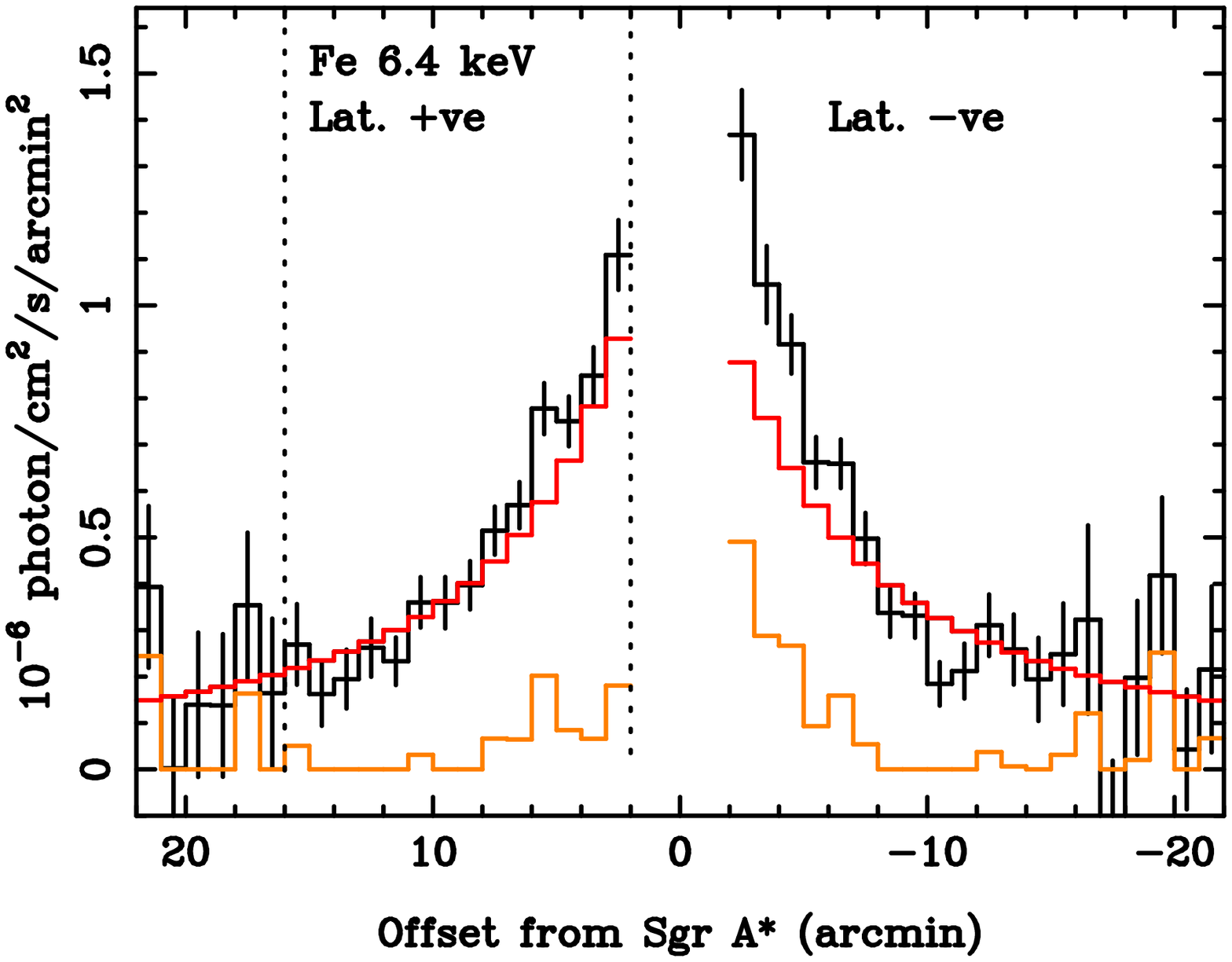}\\
\includegraphics[width=45mm, angle=270]{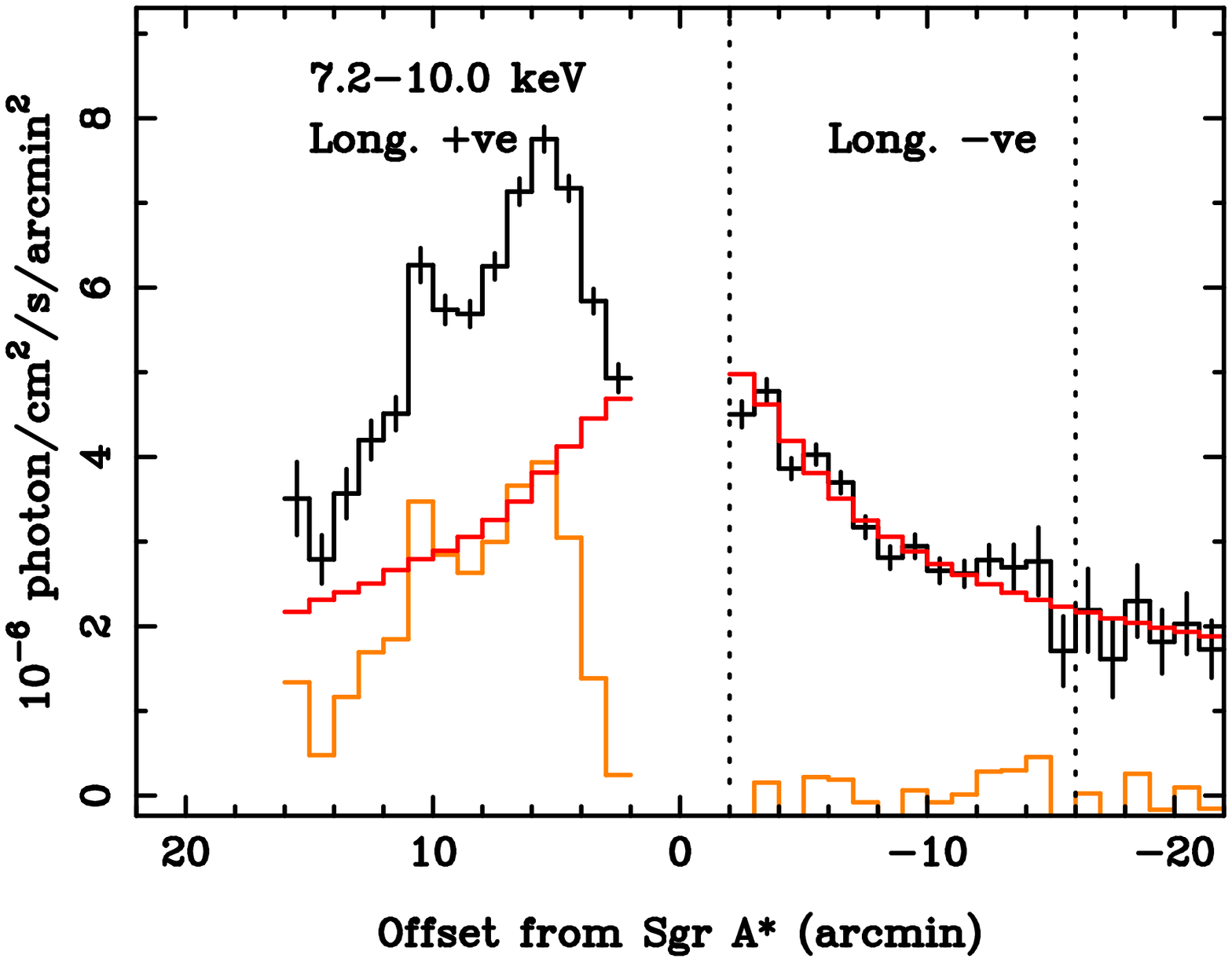}&
\includegraphics[width=45mm, angle=270]{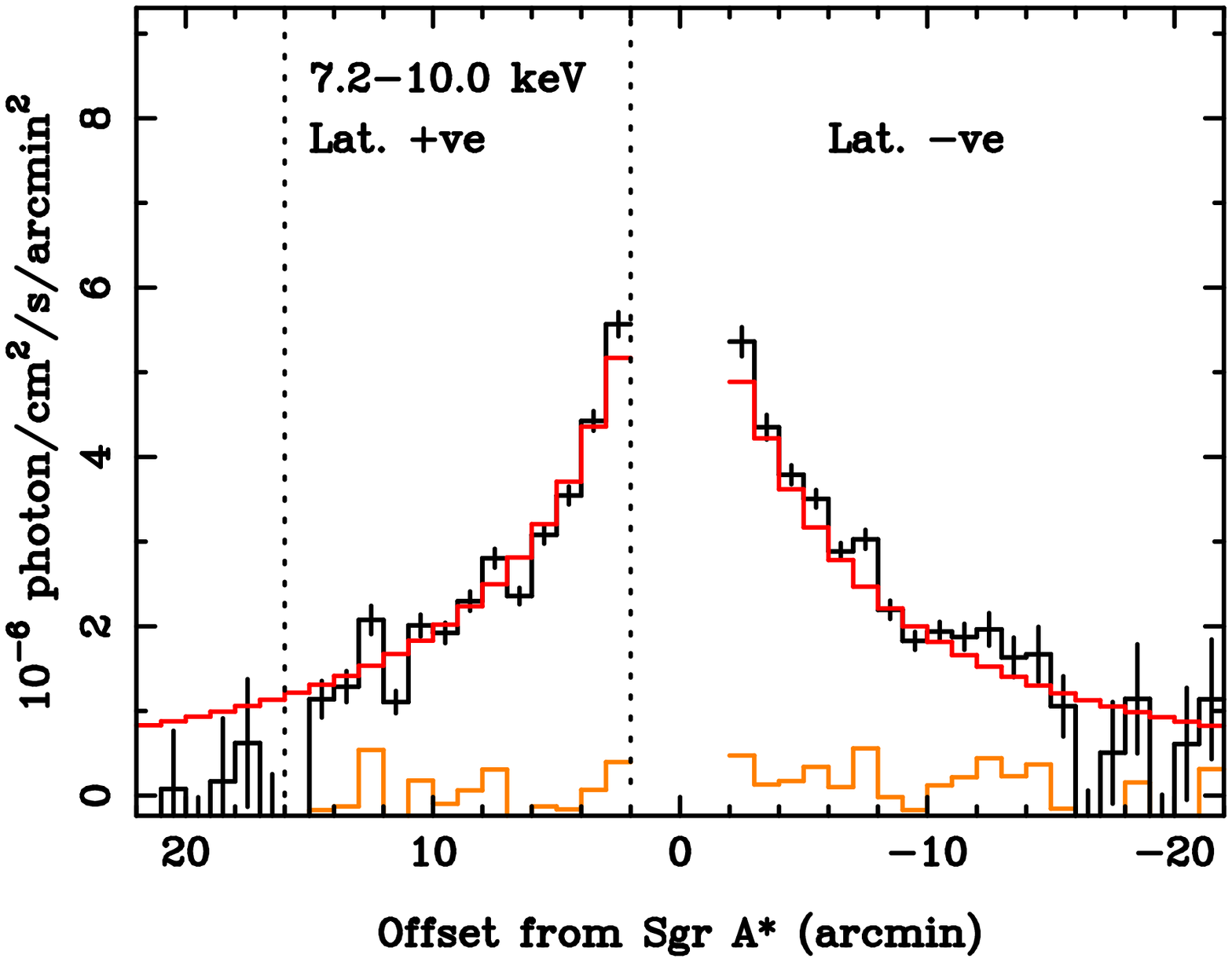}\\
\\
\end{tabular}
\caption{Spatial cuts through the GC images from the pn camera.
The left-hand panels show the longitudinal cuts (where +ve offset is to
the Galactic east of Sgr A*) and the right-hand panels show the latitudinal
cuts (where +ve offset is to the Galactic north of Sgr A*).
The spatial resolution is $1\arcmin$ along the cut direction. 
The cuts have a full width of $8\arcmin$ in the perpendicular direction.
The black points and  histograms represent the data.
The best-fitting unresolved source model is shown as the red histogram.
The data segments used to fit the source component are identified by
the vertical dotted lines.
{\it Upper panels:} Cuts through the 6.7-keV iron-line image.
The blue histogram represents the residual 6.7-keV emission once the
source component is subtracted (only +ve residuals are considered);
this excess signal may be associated with the $kT \approx 1.5$ keV
thermal emission component.
{\it Upper-mid panels:} Cuts through the 6.9-keV iron-line image.
{\it Lower-mid panels:} Cuts through the 6.4-keV iron-line image.
The orange histogram represents the residual 6.4-keV emission once 
the contribution of the unresolved-sources is subtracted (only +ve residuals
are considered); this excess signal is associated with the fluorescence of
dense molecular clouds.
{\it Lower panels:} Cuts through the 7.2--10 keV band image.  
The orange histogram represents the residual emission once the contribution
of the unresolved-sources is subtracted (only +ve residuals are considered); 
this excess signal is associated with reflection from dense molecular
clouds.
}
\label{fig:fedists}
\end{figure*}

\subsection{Spatial distribution of the unresolved sources}
\label{sec:sources}

As a means of investigating the spatial distribution of the unresolved
source population, we have employed an empirical 2-d surface brightness model
which can be matched to the available spatial information.
We adopt a model in which the surface brightness, $S$, decreases as
a power-law function of the angular offset  $\theta$ (measured
in arcmin) from Sgr A*.  The observations demonstrate very clearly that
the rate of decrease is faster perpendicular to the Galactic Plane than along the plane. We accommodate this by including an exponential fall-off 
with the latitudinal angle $\phi$, where  $\phi$ is measured 
with respect to the  Galactic latitude of Sgr A*.
The resulting model was:

\begin{equation}
\label{eq:smodel}
S = N {\theta}^{-\alpha} exp(-\frac{|\phi|}{\phi_{sc}}) +  S_{fg},
\end{equation}

\noindent where the parameters $N$, $\alpha$, and  $\phi_{sc}$, represent 
the normalisation of the power-law component, the  power-law index, and
latitudinal scale height (in arcmin), respectively.
The surface brightness is presumed to peak at the position of Sgr A*, although
in the model we actually constrained the power-law component interior 
to $\theta = 1\arcmin$ to the value at $1\arcmin$.
The model also includes a constant foreground surface brightness
$S_{fg}$. We set the latter equal to 10 per cent of the value of the variable
component at $\theta=12\arcmin$, $\phi=0\arcmin$.  
The 10 per cent factor matches the decrease in surface brightness in the
6.7-keV Fe \textsc{xxv} K${\alpha}$ line measured by \textit{Suzaku}
as the offset from Sgr A*
increases from 12\arcmin~to 1\deg \citep{uchiyama11}.
In the present analysis $S_{fg}$ represents the contribution of
sources outside of the central few hundred parsec, \textit{i.e.,} 
sources located either in the Galactic bulge or Galactic disc
(\citealt{hong09}; \citealt{uchiyama11}). 
  
\subsection{Modelling the spatial cuts}
\label{sec:cuts2}

Longitudinal and latitudinal spatial cuts derived from the 6.7-keV,
6.9-keV, and 6.4-keV iron-line images and the very-hard 7.2--10 keV 
continuum image
are shown in Fig. \ref{fig:fedists} (in all cases these are measurements
from the pn camera).  Hereafter, we use the shorthand:  Long +ve; Long -ve;
Lat +ve; and Lat -ve to identify the different segments of these cuts, as per
the labelling in Fig. \ref{fig:fedists}.

A rapid decline in the surface brightness as the distance from Sgr A* increases
is evident in most cases (with the notable exception of
the 6.4 keV and very-hard continuum measurements in the Long +ve region),
with the rate of decline being somewhat faster in the latitudinal direction. 
In order to match our empirical 2-d surface brightness model to these data, 
we derived a set of model distributions in the format of the 1-d cuts,
corresponding to an $\alpha$ versus $\phi_{sc}$ parameter grid.
Here the process was: (i) construct a model image in our standard format
for the specified $\alpha$ and $\phi_{sc}$; (ii) apply the source spatial mask,
thereby mimicking the filtering of the actual data; and (iii) determine the 
spatial cut distribution (as per \S\ref{sec:cuts}).

\begin{figure}
\centering                                 
\includegraphics[width=65mm, angle=270]{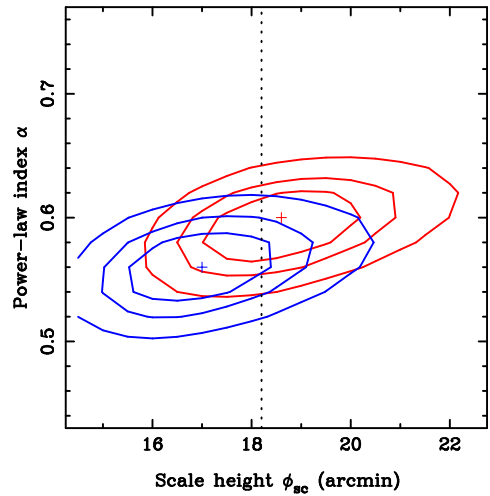}
\caption{The variation of $\chi$$^{2}$ as a function of the source
model parameters, $\alpha$ and $\phi_{sc}$. The contours  represent the
68,  90,  and  99 per  cent  confidence intervals.  The results from
the pn data are shown in red and those from the combined-MOS
data in blue. The dashed vertical line is drawn at $18.2\arcmin$;
this corresponds to the scale height of 45 pc reported for 
the Nuclear Stellar Disc by \citet{launhardt02} (see text).}
\label{fig:contours}
\end{figure}

We then selected the 6.7-keV and 6.9-keV line data pertaining to
the two ``cleanest'' regions, namely the Long -ve and Lat +ve segments
in the offset range 2\arcmin--16\arcmin, for fitting.
The fitting process involved the determination of
four free parameters, $\alpha$, $\phi_{sc}$ and two normalisations
(one for each line) via $\chi$$^{2}$-minimisation.
The results for the two ``interesting'' parameters,  $\alpha$ and 
$\phi_{sc}$ are shown in the form of a 2-d confidence
contour plot in Fig. \ref{fig:contours}.
The best-fit values were $\alpha$=0.60$^{+0.02}_{-0.03}$
and $\phi_{sc}$=18.6$^{+1.6}_{-1.2}$ arcmin, respectively.
The $\chi$$^{2}$ for the best-fit
was 109.0 for 54 degrees of freedom (dof), reflecting the fact that
there is some (modest) scatter over and above the
predicted level of the noise. Nevertheless, 
the best-fit model does account for the bulk of the 6.7-keV and 6.9-keV line
emission in the Long -ve and Lat +ve regions (see Fig. \ref{fig:fedists}).
As a check we have repeated exactly the same
analysis for the combined-MOS data.  The derived constraints
on the two model parameters $\alpha$ and $\phi_{sc}$ from the
MOS are also shown in Fig. \ref{fig:contours}; the MOS results
($\alpha$=0.56$^{+0.03}_{-0.02}$
and $\phi_{sc}$=17.0$^{+1.4}_{-1.4}$ arcmin;
$\chi$$^{2}$ = 79.6 for 54  dof)
are fully consistent with those obtained from the pn channel,
indicating that our modelling process is robust.

As a further test, using the
best-fit values for $\alpha$ and $\phi_{sc}$ derived from the pn
data, have also we fitted the Long -ve and Lat +ve cuts (2\arcmin--16\arcmin)
for the 6.4-keV iron line and very-hard band data, which
entailed determining the best-fit model normalisation for the respective
bands\footnote{In the case of the 7.2--10 keV band, we obtained
our best result when we subtracted a constant offset from the cut
data of $7.1 \times 10^{-7}$ photon/cm$^{2}$/s/arcmin$^{2}$ - this may reflect the 
limitations of the background subtraction in this continuum band.}.
The results, shown in Fig. \ref{fig:fedists}, indicate
that that the unresolved source model also accounts for the bulk of
the emission measured in these bands within the fitted regions.

Finally the models derived for the Long -ve and Lat +ve regions
were also extrapolated to the Long +ve and Lat -ve regions (with
all the parameters including the normalisations fixed). 
Fig. \ref{fig:fedists} shows the results.  In the Long +ve segment,
the unresolved source model generally under-predicts the observed signal.
This is perhaps not surprising given that this is the region to
the north-east of Sgr A*, where X-ray fluorescence
and soft thermal emission are very
prominent (Fig. \ref{fig:ims} \& \ref{fig:lineims}). However, 
the model does provide quite a reasonable description of the 
6.9-keV line emission in this region. 

In the Lat -ve segment, there appears to be an enhancement 
in  6.7-keV line emission
over and above the source model contribution. This excess extends out to an
offset of $\sim 10\arcmin$ ($\sim 25$ pc). There is some evidence for
excess 6.4-keV emission in this same region, but in this case extending
only to $\sim 5\arcmin$.  Interestingly, \textit{Chandra} data provide
evidence of some unusual emission clumps
to the south of Sgr A*, which arguably form part of a larger-scale
bipolar feature (\citealt{baganoff03}; \citealt{morris03}).
Plausibly, these are all elements of an extended structure linking 
back to the Sgr A East region
(Heard \& Warwick, 2012 , {\it in preparation}).  Nevertheless, the 
source model again provides a good match to the 6.9-keV and
very-hard band cuts in the Lat -ve region.

In summary, our modelling of the GC spatial cuts leads us to the
conclusion that the X-ray emission observed in the 6--10 keV
bandpass (encompassing three prominent iron lines Fe \textsc{i} K$\alpha$,
Fe \textsc{xxv} K$\alpha$, and Fe \textsc{xxvi} Ly$\alpha$ at 6.4, 6.7, and 6.9
keV respectively plus a very-hard continuum), to a large extent
follows a smooth, symmetric, centrally-concentrated spatial
distribution consistent
with an origin in unresolved point sources.  There are two regions
where additional emission components are clearly required. 
In the Long +ve region, 
X-ray fluorescence and reflection from dense molecular clouds is a major
contributor to the 6.4-keV line emission and the very-hard continuum band.
In this same region, the excess 6.7-keV line flux may be interpreted
in terms of soft-thermal emission, with the lack of an associated excess
at 6.9-keV constraining the temperature of the plasma to be less than
a few keV. Excess 6.7 keV emission is also seen in the Lat -ve region, 
in a structure which may traced back to the central 10 pc region.

The next step was to investigate whether the spectral characteristics
of the centrally-concentrated  component are consistent with
a putative origin in the integrated emission of low-luminosity sources.
An indication of the spectral characteristics of this component is provided
by the source model normalisations determined for
each of the bands considered in Fig. \ref{fig:fedists}. 
The normalisation of the source model for the 6.7-keV band 
relative to that measured in the 7.2--10 keV band implies
an equivalent width for the 6.7-keV iron
line of $\approx 630$ eV (assuming the underlying continuum has a
7.5-keV thermal bremsstrahlung spectral form).
Similarly, the source model normalisations measured for the 6.4-keV and 6.9-keV bands imply equivalent widths in excess of 200 eV for both lines.
This characterisation of the source emission in terms of a hard
continuum coupled with a trio of prominent iron lines is
substantiated by the results of spectral fitting presented 
in the next section.

\subsection{Modelling the X-ray spectra}
\label{sec:spec2}

\begin{figure*}
\centering
\begin{tabular}{cc}
\includegraphics[width=60mm, angle=270]{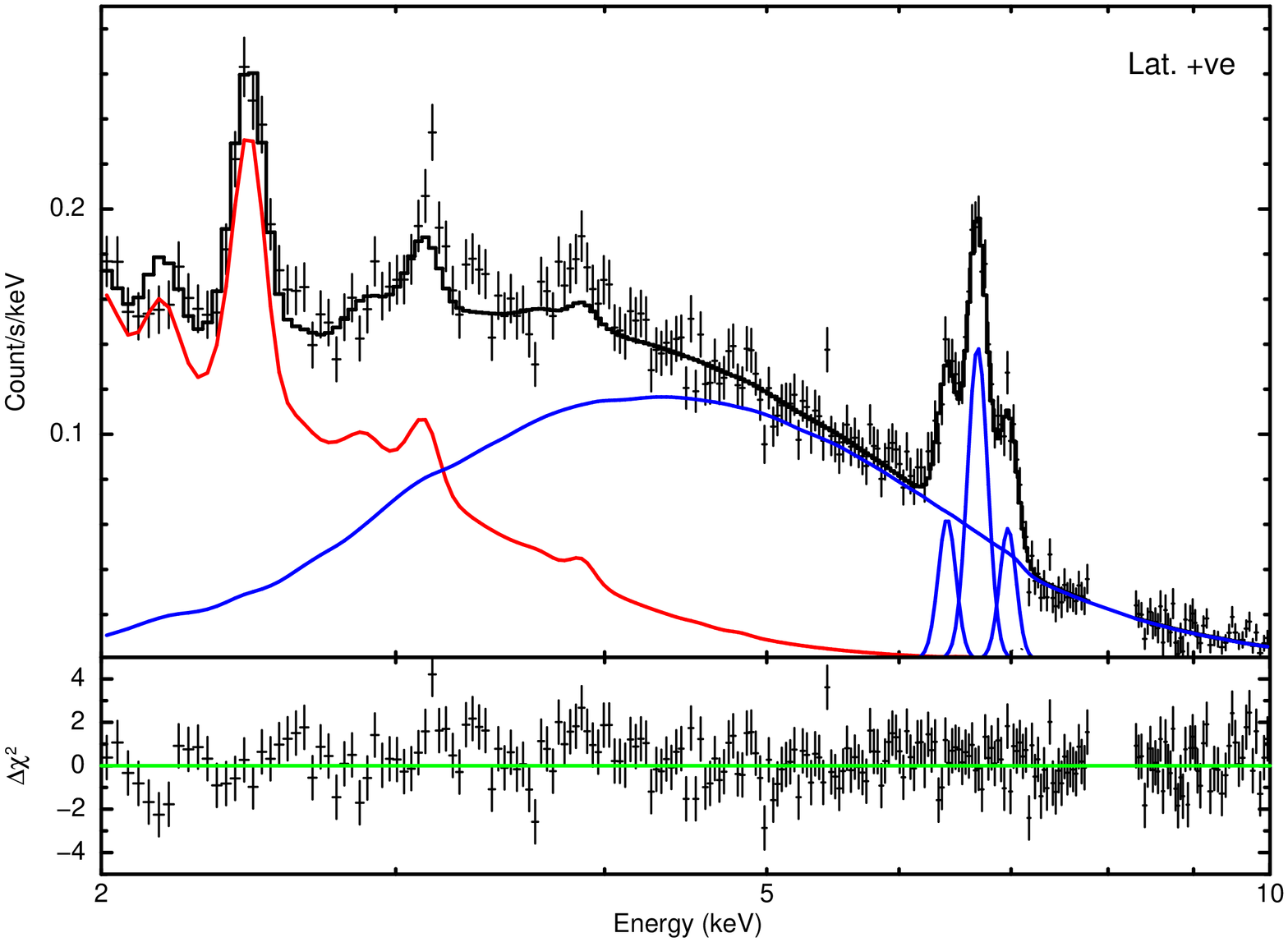}&
\includegraphics[width=60mm, angle=270]{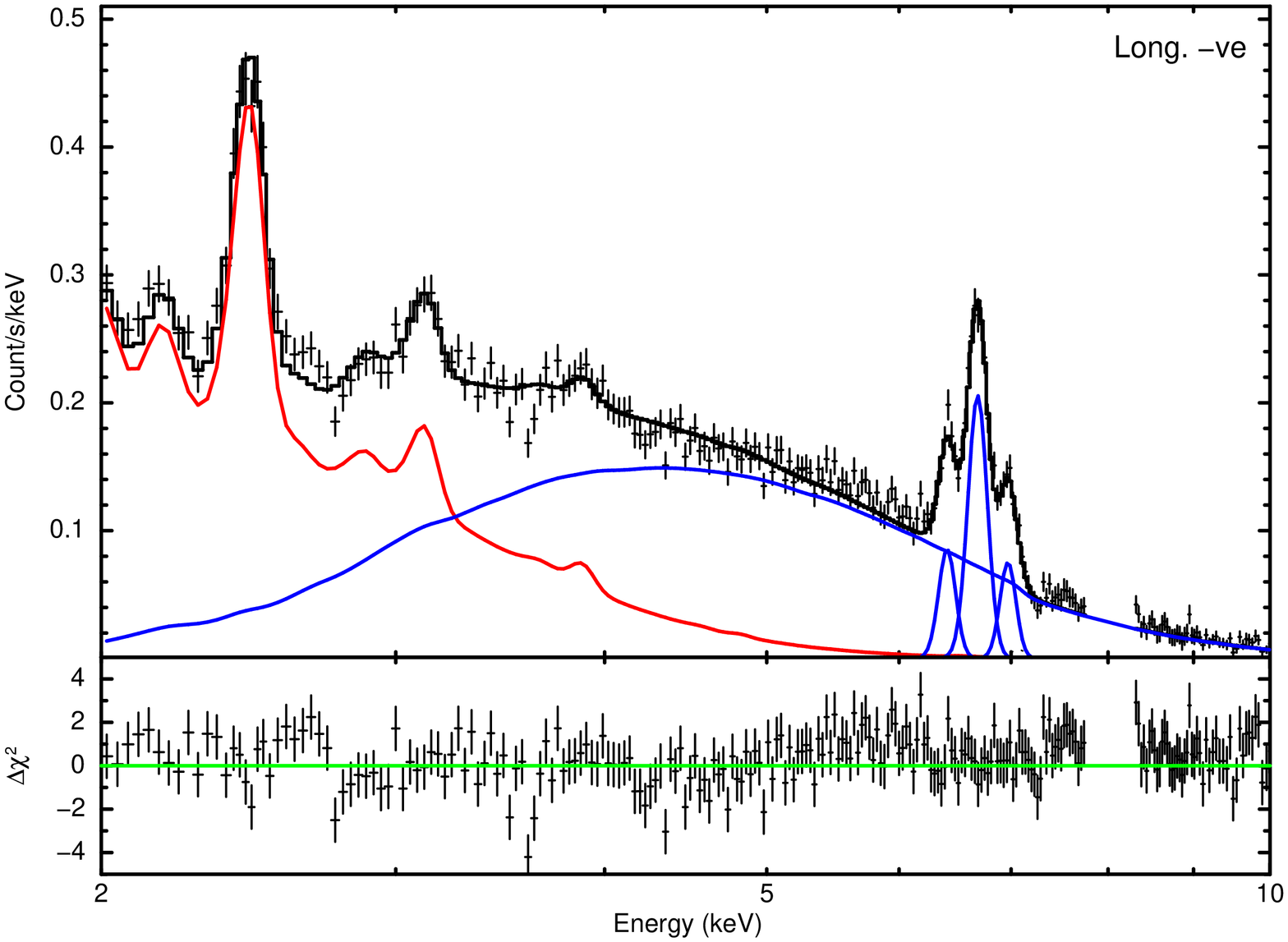}\\
\includegraphics[width=60mm, angle=270]{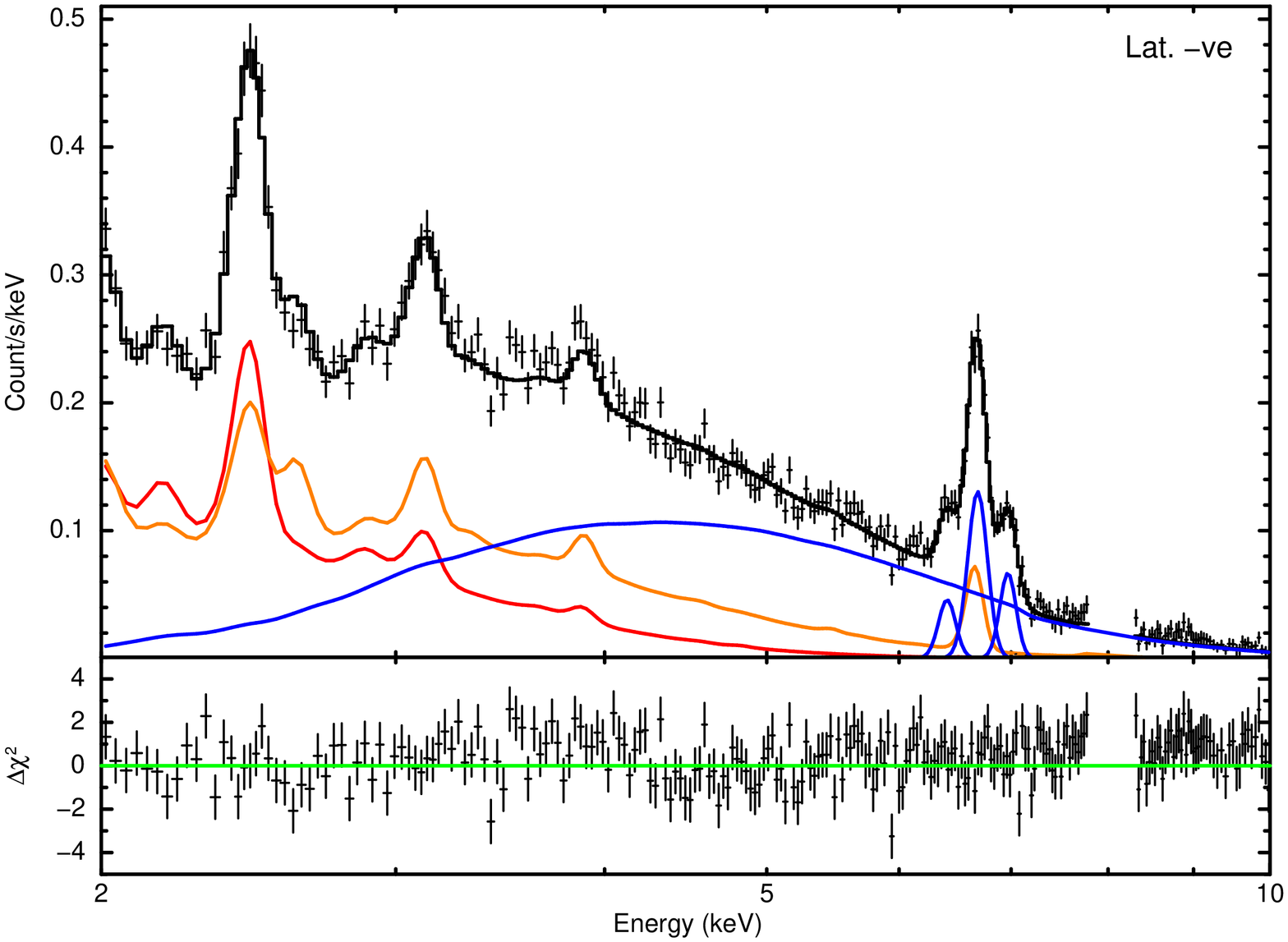}&
\includegraphics[width=60mm, angle=270]{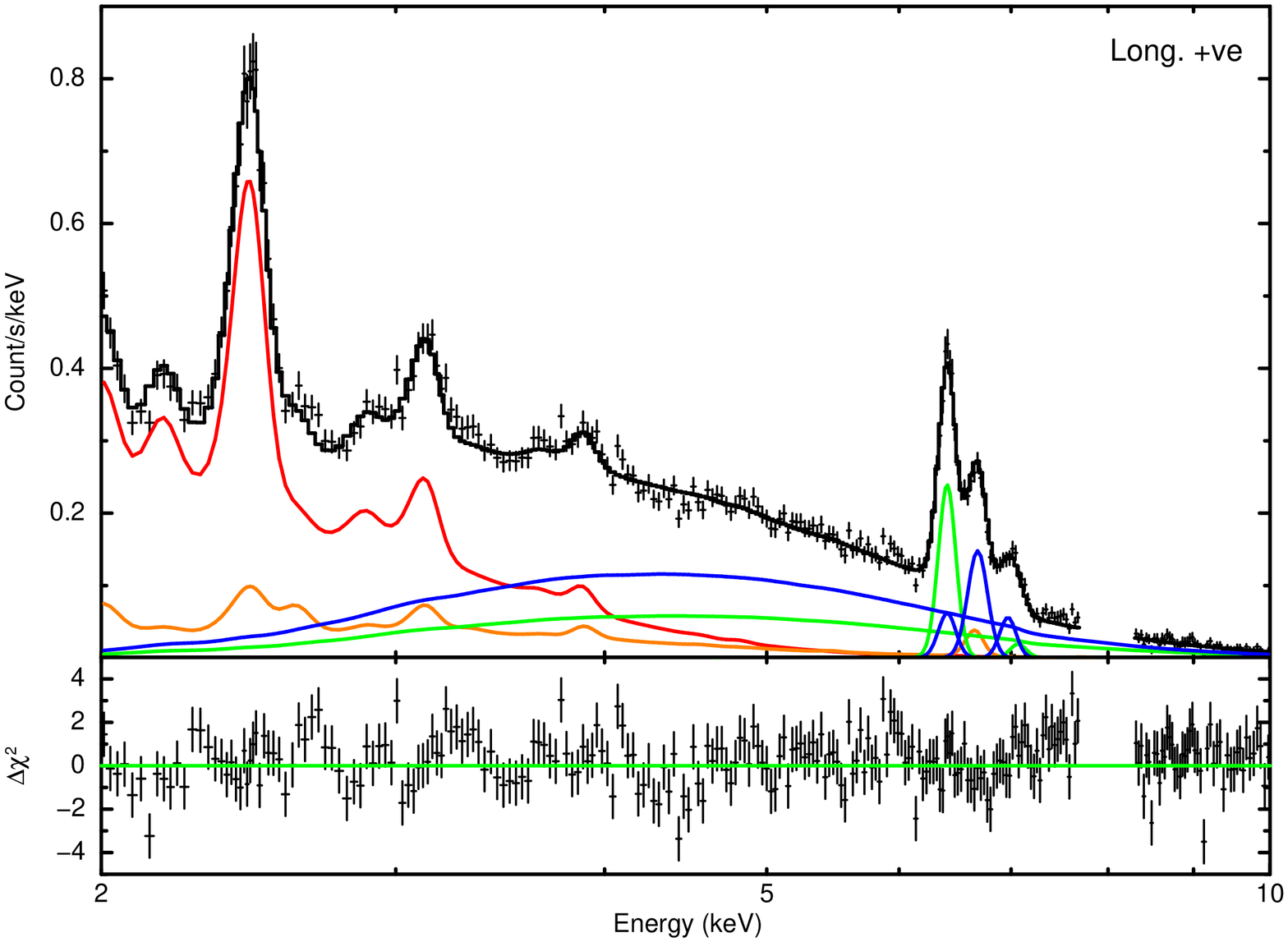}\\
\end{tabular}
\caption{X-ray spectra measured by the pn camera in observation
0202670801 in four regions - {\it Upper-left  panel:} 
the Lat +ve region; {\it Upper-right  panel:} the Long -ve region;
{\it Lower-left  panel:} the Lat -ve region; 
{\it Lower-right  panel:} the Long +ve region.
 Each panel shows the X-ray spectrum
after background subtraction in the 2--10 keV band, excluding the
region around the Cu K$\alpha$ instrumental line. The best-fitting
model and the fitting residuals are also shown. 
The contribution of different spectral components is illustrated
by the different colour curves - {\it blue:} the 7.5  keV thermal
bremsstrahlung continuum plus iron lines associated with
unresolved sources;
{\it green:} the fluorescent iron line and reflected continuum
from dense molecular clouds; {\it red:} soft thermal emission 
with $kT = 0.8$ keV;
{\it orange:} (relatively) soft thermal emission with $kT = 1.5$ keV.}
\label{fig:brspec}
\end{figure*}

The X-ray spectra measured by the pn camera in the four regions identified 
in Fig. \ref{fig:ims}  (which we again refer to as the Long +ve, Long -ve, 
Lat +ve, Lat -ve regions) are shown in Fig. \ref{fig:brspec}. In the spectral
fitting of these data we have considered the three emission components
discussed earlier ($\S$\ref{sec:distinct}). 

The first component represents the integrated emission from unresolved
sources and was characterised in the spectral fitting as a
\texttt{bremss} continuum  with $kT = 7.5$ keV, plus Gaussian-line 
components corresponding to the Fe \textsc{i} K$\alpha$,
Fe \textsc{xxv} K$\alpha$, and Fe \textsc{xxvi} Ly$\alpha$ lines.
The intrinsic width of the 6.4-keV and 6.9-keV lines was fixed at
30 eV, in line with previous estimates of apparent line-widths
in pn spectra \citep{capelli12}. In the case of
the 6.7-keV He-like iron line, preliminary fitting of  the Lat +ve spectrum
indicated a best-fit value  for the intrinsic width of $\sim55$ eV, 
and in the subsequent spectral analysis the intrinsic width of this line was
fixed at this value. Similarly the energies of the three iron lines were 
measured to be  6.41 $\pm$ 0.01 keV, 6.68 $\pm$ 0.01 keV, and 6.97 $\pm$ 0.01
keV, and subsequently fixed at these values. An Fe \textsc{i} K$\beta$ line
was included at an energy of 7.07 keV and width 30 eV with a
normalisation tied to 11 per cent of that of the
Fe \textsc{i} K$\alpha$ line \citep{koyama09}.  

The second emission component representing the fluorescence and reflection from
dense molecular clouds was modelled as a power-law continuum with 
photon index $\Gamma$ plus two Gaussian lines corresponding 
to neutral Fe K$\alpha$ and  K$\beta$ (with parameters other than the
normalisation of the Fe K$\alpha$ line set as above).   

Soft thermal emission constituted the final ingredient of the composite
spectral model. Twin \texttt{apec} components with temperatures fixed
at $kT_{1}=0.8$ keV and $kT_{2} = 1.5$ keV were used to represent the
likely complex temperature structure. In the spectral fitting the
metal abundance was allowed to vary from region to region, 
but in all cases was tied across the two thermal components.

It was assumed that each of above components was subject to absorption. 
For the soft thermal emission we set $N_{\rm{H}} = 6.5 \times
10^{22}$ cm$^{-2}$, in line with recent estimates
of the soft X-ray  absorption along the line-of-sight to the GC
(e.g., \citealt{muno06}).
In the case of the integrated emission of the unresolved
sources and the  reflection/scattering from molecular clouds, the absorption
column density was increased to  $N_{\rm{H}} = 12 \times 10^{22}$ cm$^{-2}$,
so as to include an additional contribution intrinsic either to the source
spectra or the molecular cloud environment. Unfortunately the intrinsic absorption
is rather poorly constrained due to the pervasive soft thermal emission, thus a value of $5.5 \times 10^{22}$ cm$^{-2}$ was eventually chosen so 
as to give reasonably consistent results across the four regions. 

With all the elements of the composite spectral model in place\footnote{
Our spectra model also included a power-law continuum with photon index $\Gamma = 1.4$,
subject to an absorbing column density $N_{\rm{H}} = 13 \times 10^{22}$ cm$^{-2}$
with a normalisation matched to that of the CXB (\citealt{lumb02}; \citealt{deluca04}). However, the inclusion of this component had negligible impact on the spectral fitting results.}, we
first attempted to fit the spectra of the Lat +ve and Long -ve regions 
with the integrated source emission model (involving four free parameters,
\textit{i.e.,} the normalisations of the bremsstrahlung continuum and the three iron emission lines) together with a  
$kT=0.8$ keV \texttt{apec} component (involving two further free parameters - the \texttt{apec} normalisation and metal abundance). This restricted model
provided a good match to the observational data
(Fig. \ref{fig:brspec}) with best-fitting parameters as listed in
Table \ref{tab:parameters}. Of particular note is the fact that the
derived equivalent widths for each line  are reasonably consistent across the
two regions, and broadly in line with the expectations based on the spatial cuts
analysis ($\S$\ref{sec:cuts2}). Based on our best-fitting unresolved source model, a 2--10 keV flux of $1.1 \times 10^{-11}$ erg s$^{-1}$ cm$^{-2}$ has, on average, an associated 6.7-keV photon flux of $4.0 \times 10^{-5}$ photon s$^{-1}$ cm$^{-2}$.

In extending the spectral analysis to the Lat -ve region, it was necessary
to introduce the hotter $kT=1.5$ keV \texttt{apec} component into the fitting
(thus accounting for the excess 6.7-keV iron line emission observed in 
this region - see \S\ref{sec:cuts2}). The best-fit model for this region
is again illustrated in 
Fig. \ref{fig:brspec} and the best-fit parameter values tabulated in 
Table \ref{tab:parameters}. Again the parameters for the
integrated source emission are strikingly similar to those of the
Lat +ve region.

 Finally, for the Long +ve region we required the full composite spectral
model, consistent with our earlier identification of this region as the most
active of the four considered. Here the spectral-fitting approach was to
fix the parameters of the unresolved-source emission model at those
obtained from the Long -ve region.  The additional free-parameters
were those quantifying the reflection from the
molecular clouds ({\it i.e.,} the continuum normalisation and photon index) and
the fluorescence ({\it i.e.,} the normalisation of the 6.4-keV line)
Again,  Fig. \ref{fig:brspec} and Table \ref{tab:parameters} summarise the
results. The derived photon index of $\Gamma \approx 1.8$ is consistent 
with other recent estimates \citep{capelli12}. 
Similarly the inferred
large equivalent width of the fluorescent line ($\sim1.7$ keV) is in line
with published values (\citealt{ponti10}; \citealt{capelli11}, \citeyear{capelli12}).

An interesting result from Table \ref{tab:parameters} is the inferred
variation in the metal abundance across the four regions, with values
ranging from 0.5--2~$Z_{\rm{\odot}}$. We measure the highest metal
abundance to the north-east of Sgr A* (Long +ve), the site with the
most recent supernova activity, which could have led to a localised
enrichment of the thermal plasma. A \textit{Suzaku} study of the
GC X-ray emission in the same region determined the abundance 
to be in the range $\sim$~1--2~$Z_{\rm{\odot}}$ \citep{nobukawa10},
consistent with the current measurements.

\begin{table*}
\begin{center}
\caption{The best-fitting spectral parameters for the four regions.}
\begin{tabular}{l c c c c}
\hline
Region & Lat +ve & Long -ve & Lat -ve & Long +ve \\ 
\hline 
\multicolumn{5}{c}{Unresolved source component}\\
\hline
$N_{\rm{H}}$ ($\times 10^{22}$ cm$^{-2}$)&12&12&12&12\\
$kT$ (keV)&7.5&7.5&7.5&7.5\\
$Norm.$ ($\times 10^{-3}$) &$5.90\pm{0.09}$& $6.93\pm{0.09}$& $5.92\pm{0.19}$&6.93\\
$EW_{\rm{6.4}}$ (eV) &$236\pm{27}$&$233\pm{23}$&$188\pm{30}$&$233$\\
$EW_{\rm{6.7}}$ (eV) &$691\pm{37}$&$781\pm{37}$&$710\pm{58}$&$781$\\
$EW_{\rm{6.9}}$ (eV) &$296\pm{33}$&$283\pm{27}$&$366\pm{37}$&$283$\\
\hline
\multicolumn{5}{c}{Reflection component}\\
\hline
$N_{\rm{H}}$ ($\times 10^{22}$ cm$^{-2}$)&-&-&-&$12$\\
$\Gamma$ &--&--&--&$1.8^{+0.2}_{-0.2}$\\
$Norm.$ ($\times 10^{-3}$) &-&-&-&$3.3^{+1.4}_{-1.0}$\\
$EW_{\rm{6.4}}$ (eV)&-&-&-&$1655\pm{80}$\\
\hline
\multicolumn{5}{c}{Soft thermal plasma component}\\
\hline
$N_{\rm{H}}$ ($\times10^{22}$ cm$^{-2}$)&6.5&6.5&6.5&6.5\\
$kT_{1}$ (keV)&0.8&0.8&0.8&0.8\\
Norm. ($\times 10^{-2}$) &9.6$^{+1.0}_{-1.0}$&11.2$^{+1.1}_{-1.1}$&5.7$^{+1.0}_{-0.8}$&8.6$^{+1.5}_{-1.5}$\\
$kT_{2}$ (keV)&-&-&1.5&1.5\\ 
Norm. ($\times 10^{-2}$)&-&-&2.7$^{+0.3}_{-0.3}$&0.81$^{+0.24}_{-0.21}$\\
$Z$ ($Z_{\rm{\odot}}$)&0.49$^{+0.11}_{-0.09}$&0.73$^{+0.12}_{-0.10}$&1.07$^{+0.17}_{-0.14}$&2.18$^{+0.56}_{-0.39}$ \\
\hline
$\chi^{2}$/$\nu$&1473.7/1422&1404.1/1414&1560.9/1423& 1569.2/1407\\
\hline
\end{tabular}
\label{tab:parameters}
\end{center}
\end{table*}

In order to test the  hypothesis that the integrated source emission is
largely due to magnetic CVs,  we have employed the spectral model
for  intermediate polars (IPs) developed   by \citet{yuasa10}. 
This IP model  has three free parameters: the white dwarf
(WD)   mass  $M_{\rm{WD}}$; the relative iron abundance  $Z_{\rm{Fe}}$, 
and a normalisation factor.  When this spectral model was substituted for
the thermal bremsstrahlung continuum and the associated lines at 6.7 keV and 6.9 keV
(with the normalisation of the 6.4-keV line retained as a free parameter since
the IP model does not include 6.4-keV iron line emission),
we obtained an excellent fit to the data with a $\chi^{2}/\nu$ value 
for the joint-fitting of the spectra for the Lat +ve and Long -ve regions of
2944.8/2849.   The best-fitting values for the white dwarf  mass and  
iron abundance were $0.49\pm{0.02}$ $M_{\rm{\odot}}$  and
$0.97\pm{0.05}$  $Z_{\rm{\odot}}$.  
This estimate of  the typical mass of a magnetic white dwarf in the GC  is, in fact, very similar
to that obtained by \citet{yuasa12} in their analysis of {\it Suzaku}
XIS and HXD/PIN spectra pertaining to a set of Galactic bulge fields. 
However, as noted by these authors this mass estimate is very likely
biased downwards by other types of CV, such as dwarf novae in quiescence, 
which have lower representative
temperatures. Also the lack of spectral information above 10 keV in the
current analysis, severely compromises the measurement of the effective
temperature of the emission ($kT > 15$ keV), 
which directly impinges on the mass estimate. Nevertheless, the
analysis does demonstrate that it is very plausible that the bulk of
the emission seen above 5 keV in the {\it XMM-Newton} spectra is
attributable to an unresolved source population dominated by magnetic CVs.

\section{Discussion}

\subsection{The low-luminosity X-ray source population at the Galactic Centre}
\label{sec:massmodel}

Our spatial and spectral analysis confirms the view that much of 
the hard X-ray emission observed within $20\arcmin$ of Sgr A* can be 
attributed to the integrated emission of unresolved low-luminosity
X-ray sources. The spectral characteristics of this component -
a hard continuum (with $kT \sim7.5$ keV when modelled as thermal bremsstrahlung) plus three iron lines
(Fe \textsc{i} K$\alpha$, Fe \textsc{xxv} K$\alpha$, and 
Fe \textsc{xxvi} Ly$\alpha$) of relatively high equivalent width
- places a tight constraint on the nature of the underlying population.
These spectral properties rule-out a significant contribution from most
classes of  coronally-active stars and binaries or from young stellar
objects \citep{muno03}. High-mass and low-mass X-ray binaries 
(HMXBs and LMXBs) 
represent the most X-ray luminous X-ray population within the Galactic
disc and bulge (\citealt{muno03}; and references therein),  but it seems 
highly unlikely that very low-luminosity
versions of such sources exist in sufficient numbers in the GC to
explain the observations, although some quiescent LMXBs do seem to have
the requisite hard spectra (e.g., \citealt{wijnands05}).

Currently, the leading candidate for the unresolved source population is
magnetic CVs  (\citealt{muno04a}; \citealt{laycock05}), which in principle
include both polars and IPs. Although the spectra of these subclasses of CV
can both be characterised as multi-temperature thermal plasmas, the plasma
temperatures of polars are lower due to enhanced cyclotron cooling 
({\it e.g.,}
\citealt{cropper98}) and it is the IPs which likely dominate the
hard ($>$ 10 keV) emission seen from the GC \citep{yuasa12} and,
correspondingly, make the largest contribution to the key iron-line tracers.
Nearby, well-studied magnetic cataclysmic variables typically have
2--10 keV X-ray luminosities in the range from $10^{31}$ -- 
$10^{34}$ erg s$^{-1}$
({\it e.g.,} \citealt{sazonov06}), although the true low-luminosity
bound for the population may be at least one, and possibly two,
orders of magnitude fainter \citep{revnivtsev09}.
A key question for the unresolved source model is
whether there are sufficient magnetic CVs
in the GC to account for the bulk of the observed hard X-ray luminosity.

\begin{figure*}
\centering
\begin{tabular}{c c}
\includegraphics[width=60mm, angle=270]{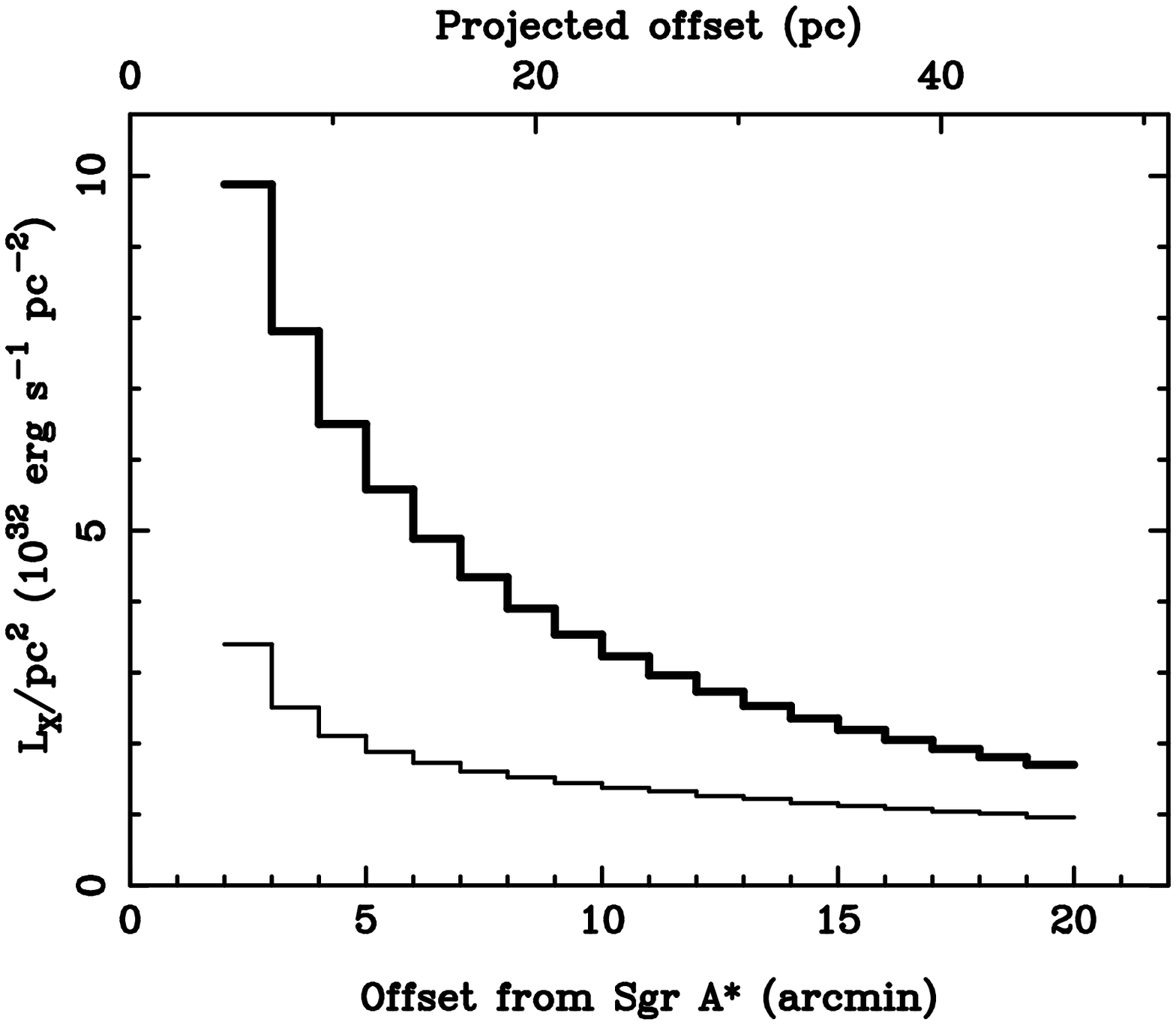}&
\includegraphics[width=60mm, angle=270]{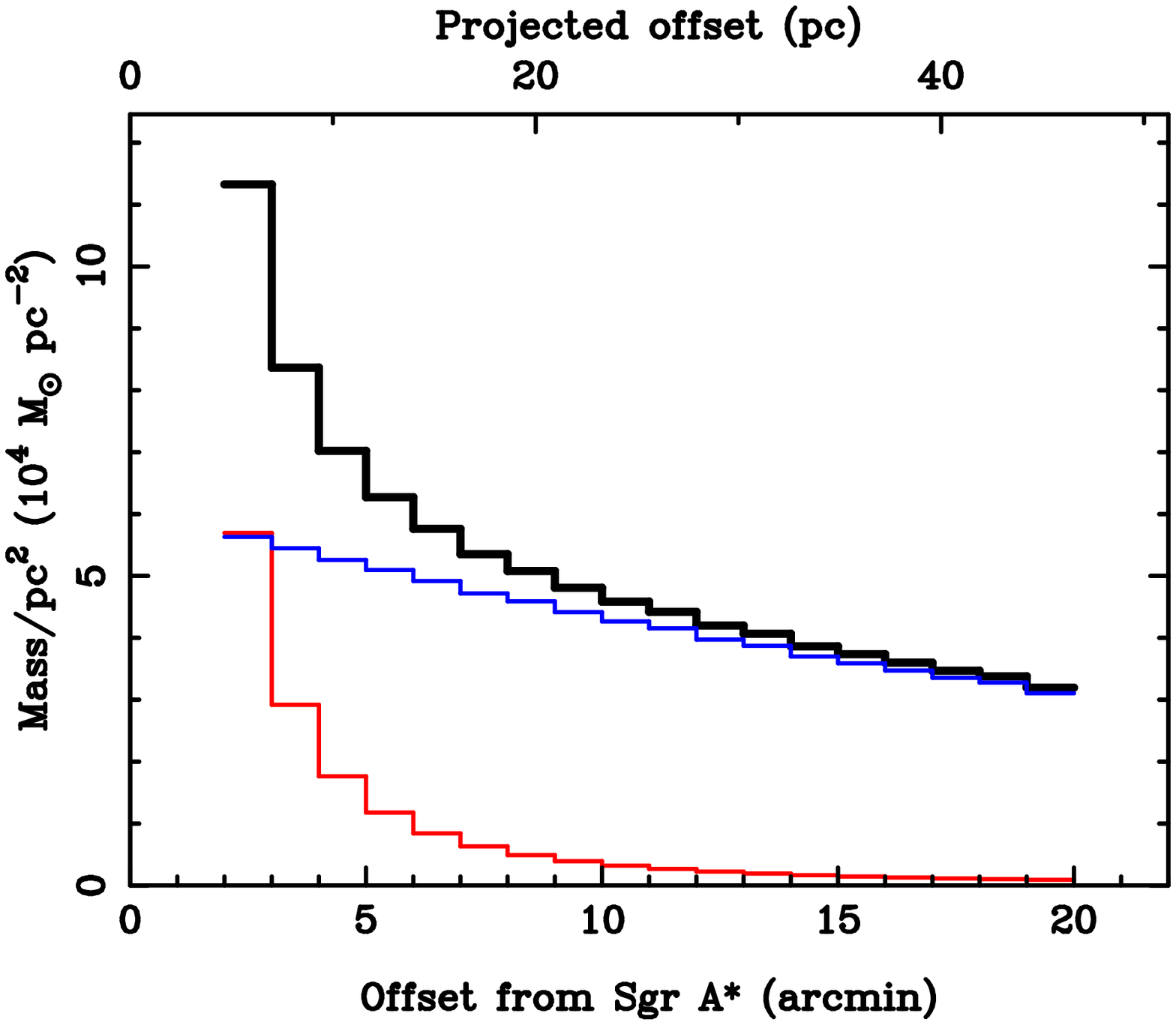}\\
\includegraphics[width=60mm, angle=270]{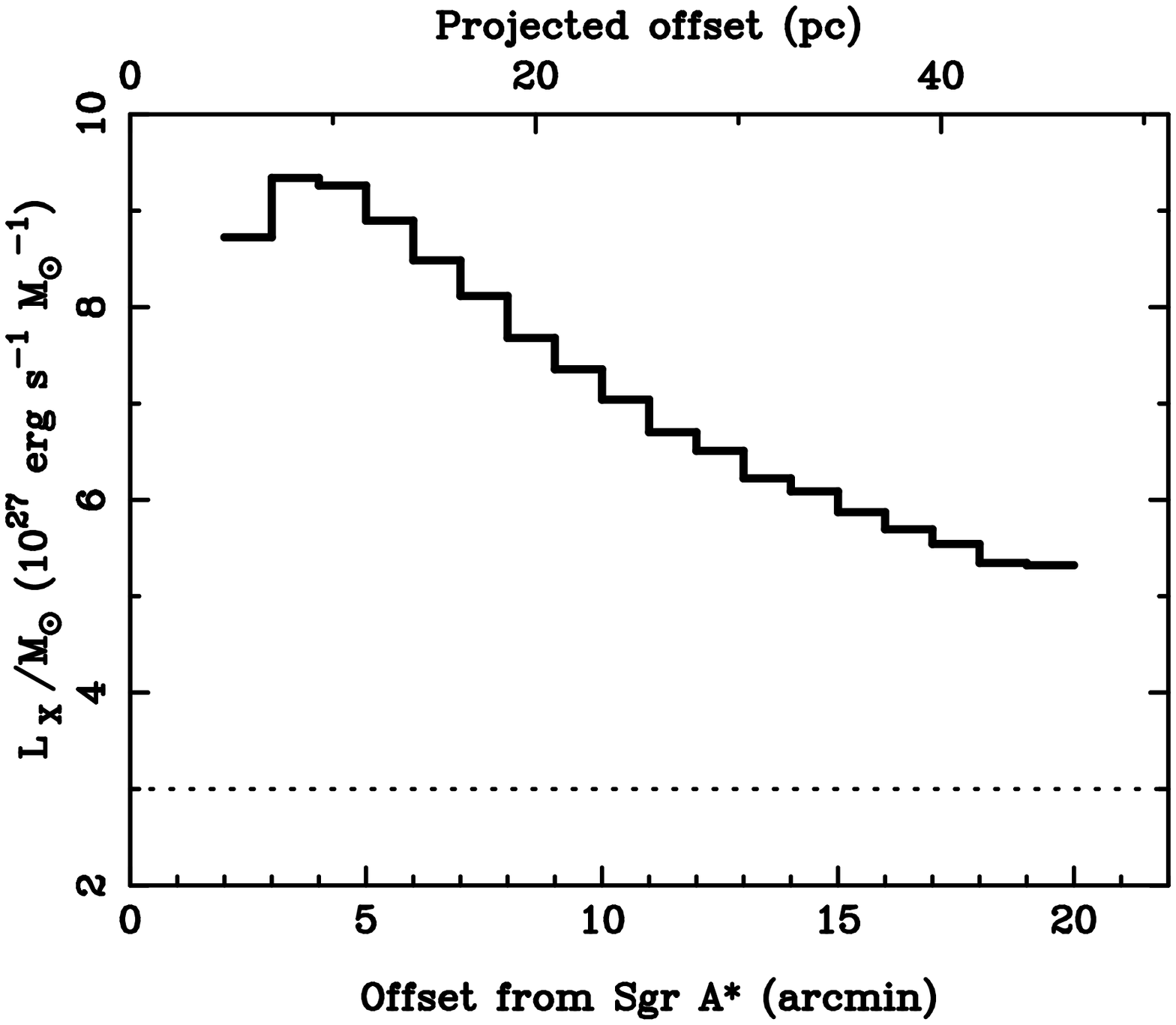}&
\includegraphics[trim=-15mm 0mm 0mm 0mm,clip,width=60mm, angle=270]{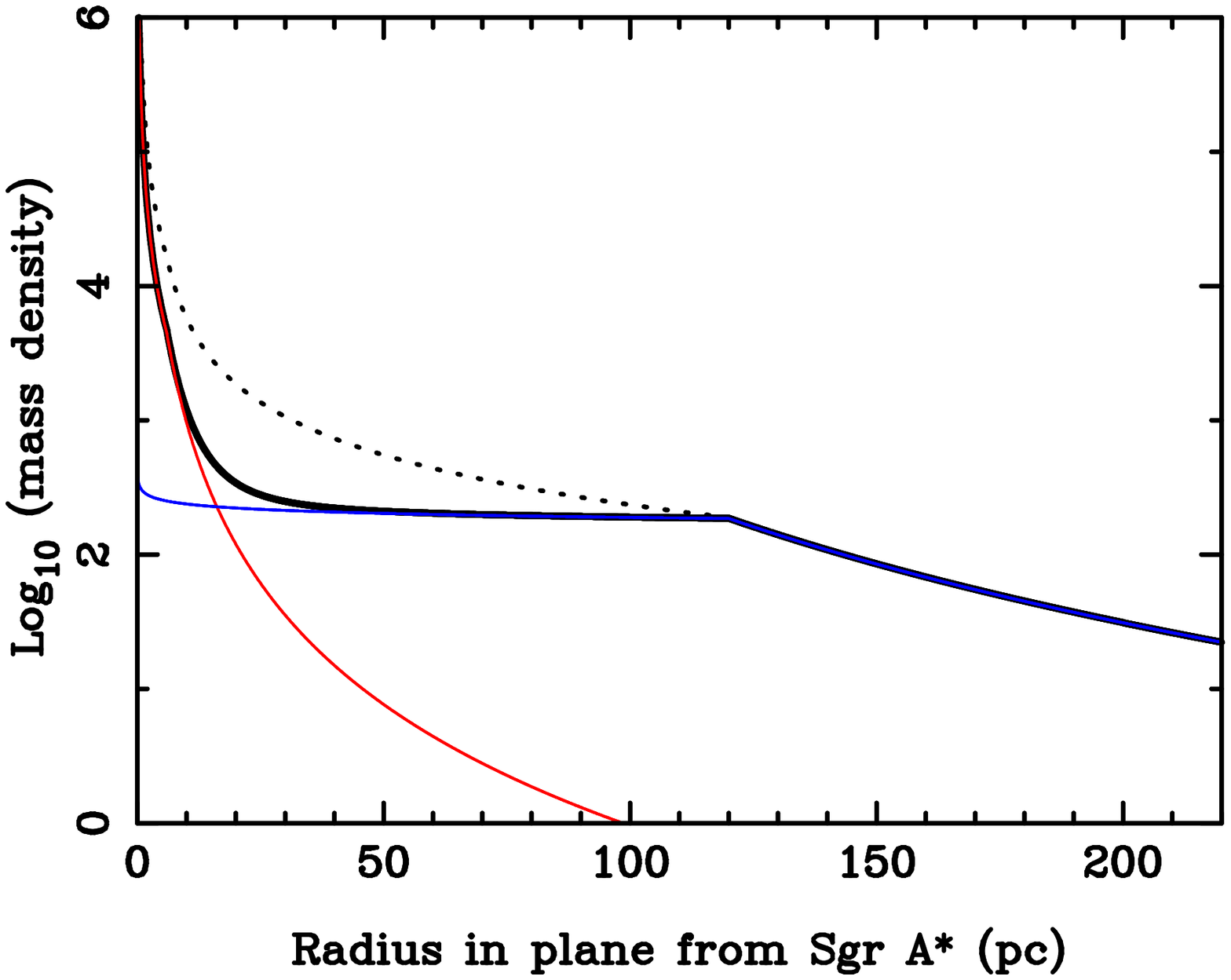}\\
\end{tabular}
\caption{{\it Upper-left panel:} Variation of the 2--10 keV X-ray surface brightness  (in units of $10^{32} \rm~erg~s^{-1}~pc^{-2}$) as a function of the angular offset from Sgr A*. The thick and thin black lines are the luminosity per square parsec derived from the X-ray images and predicted from the mass-model respectively. {\it Upper-right panel:} Projected mass density (in units of $10^{4}~\Msun~\rm pc^{-2}$) in the GC showing the contribution of the NSC (red), the NSD (blue), and their sum (black). {\it Lower-left panel:} The inferred X-ray emissivity per unit stellar mass (in units of $10^{27} \rm~erg~s^{-1}~\Msun^{-1}$) as a function of the angular offset from Sgr A*. The dotted line is the  ``base-level'' X-ray emissivity which is representative of measured GRXE values. {\it Lower-right panel:} The variation in the stellar mass density on the plane (in units of $\Msun\rm~pc^{-3}$) for the GC region as a function of the distance (in pc) from Sgr A*. The solid black curve represents the sum of the NSC (red) and the NSD (blue) components in the \citet{muno06} model. The dotted black curve is the mass-model (sum of the NSC and NSD components) inferred on the basis of the assumption that the X-ray emissivity in the region is $3 \times 10^{27} \rm~erg~s^{-1}~\Msun^{-1}$.}
\label{fig:diff}
\end{figure*}

As a starting point, we have determined the integrated
X-ray luminosity, $L_{\rm{X}}$,
of the unresolved source component.  For this purpose, we first measured
the normalisation of the empirical surface brightness model,
when fitted to the Long -ve and Lat +ve segments of the spatial cuts pertaining to the 4.5--6 keV pn images (noting that in these regions, the unresolved
sources appear to provide the dominant contribution to the 4.5--6 keV flux).
Next we converted  pn count s$^{-1}$ in the 
4.5--6 keV band to X-ray flux in the full 2--10 keV band 
(via the spectral model of the unresolved source component
discussed previously), and applied the  necessary scaling factors
to transform the surface brightness model units
to X-ray luminosity per square parsec. Finally, we determined
the radial surface-brightness distribution from an image
derived from the model\footnote{We use circular annuli of width
$1\arcmin$. Hence the derived radial
distribution represents an average over latitude and longitude.}. 
The result is shown in the upper-left panel of Fig.\ref{fig:diff}.
Integrating this distribution from 2\arcmin--20\arcmin~ (which corresponds to a projected offset of 4.6--46 pc at the GC), we obtain $L_{\rm{X}} \approx 2.0 \times 10^{36}$ erg s$^{-1}$ as the total integrated
(2--10 keV) X-ray luminosity of the unresolved sources (below a source detection threshold of $L_{\rm{X}} \approx 10^{33}$ erg s$^{-1}$).

The next step was to consider the GC stellar mass contained within this
region. Here we use the same mass-model prescription as \citet{muno06},
who in turn use parameters obtained from an analysis of
\textit{NIR} maps by \citet{launhardt02}\footnote{\citet{muno06} adopt the same parameters values as \citet{launhardt02} despite the fact that the original analysis was based on a GC distance of 8.5 kpc. However, as noted by \citet{launhardt02}, in practice, such scaling corrections are typically much smaller than the uncertainties in the parameter determinations. In the present paper we employ the same mass-model formulation as \citet{muno06}, along with the assumption of a GC distance of 8 kpc.}

Within the central 
few hundred parsecs of the Galaxy, two major stellar distributions
have been identified:  the Nuclear Stellar Cluster (NSC) and
the Nuclear  Stellar Disc (NSD).
  
The NSC has a mass density, which varies with radius $R$
as:

\begin{equation}
\label{eq:central}
\rho = \frac{\rho_{\rm{c}}}{1+(R/R_{\rm{c}})^{n_1}}.
\end{equation}

\noindent     For     $R<6$    pc,     $\rho_{\rm{c}}=3.3\times10^{6}$
$\Msun$pc$^{-3}$, $R_{\rm{c}}=0.22$ pc,  and $n_{1}=2$.  For $6<R<200$
pc, $n_{1}=3$, $R_{\rm{c}}$ remains  the same and $\rho_{\rm{c}}$ must
be adjusted so that the profile is continuous at $R=6$ pc. 

The mass distribution of the NSD can be modelled as a power-law 
plus exponential function,

\begin{equation}
\label{eq:disc}
\rho = \rho_{\rm{d}}R^{-n_2}\exp(-|z|/z_{\rm{d}}).
\end{equation}

\noindent   For  $R<120$  pc,   $\rho_{\rm{d}}=300$  $\Msun$pc$^{-3}$,
$n_{2}=0.1$,   and    $z_{\rm{d}}=45$   pc.    For    $120<R<220$   pc,
becomes $n_{2}=3.5$ with $z_{\rm{d}}$ unchanged; $\rho_{\rm{d}}$ must
be adjusted so  that the function  is continuous for $n_{2}=3.5$. 
Beyond 220 pc this disc component is assumed to
cut-off rapidly.

The other components of the mass-model discussed by \citet{muno06}
are the Galactic bulge and disc. However, these are not considered
here, since they contribute only to the constant term in our
empirical surface brightness model (on the angular scales considered),
which was not included in the derivation of the integrated luminosity.

We integrated the mass distributions in the NSC and NSD components
along a set of lines-of-sight to give the stellar mass per square parsec on a 2-d
grid of positions centred on Sgr A*. From this 2-d distribution
we then determined the average stellar-mass density within circular annuli
of width $1\arcmin$ as a function of radial offset in the range
2\arcmin--20\arcmin.
The result is shown in the upper-right panel of Fig.\ref{fig:diff},
which displays both the individual contributions of the NSC
and NSD
and their sum. It is evident that the NSD provides the dominant 
mass contribution for offset angles greater than a few arcmin
and that the NSC component becomes insignificant outside $\sim$10\arcmin.

The scale height for the NSD of $45$ pc  reported by \citet{launhardt02}, based on IRAS 100 \micron~dust measurements, corresponds to 18.2$\arcmin$ at the distance  of the GC assumed by these authors. This is in excellent agreement with the scale height inferred from the X-ray observations (see Fig. \ref{fig:contours}) and consistent with the fact that the NSD is the major contributor to the stellar mass within the region sampled by the X-ray observations.

The X-ray emissivity per unit stellar mass can be readily
determined from the ratio of the curves presented in
the upper-left and upper-right (total mass) panels of
Fig. \ref{fig:diff}.
The result is shown in the lower-left panel of Fig. \ref{fig:diff}.
A smooth trend is apparent with offset angle, with the inferred
emissivity varying from $\approx 5 \times 10^{27}$ erg s$^{-1}~\Msun^{-1}$
at $20\arcmin$ up to almost twice this value at $2\arcmin$.

By way of comparison \citet{sazonov06} constructed an X-ray luminosity function for faint sources in the local neighbourhood, from which they determined the total X-ray emissivity for CVs and active binaries (that is sources associated with the old stellar population) to be $\sim (3.1\pm0.8) \times 10^{27}$ erg s$^{-1}~\Msun^{-1}$ in the 2--10 keV band, with CVs contributing roughly one-third of this total. More recently, using a deep \textit{RXTE}/PCA scan across the Galactic Plane at $l_{\textsc{ii}} \sim +18.0\deg$, \citet{revnivtsev12} calculate the 2--10 keV emissivity in the GRXE to be $(3.0\pm0.3) \times 10^{27}$ erg s$^{-1}~\Msun^{-1}$. Moving closer to the GC, from a \textit{Chandra} deep-field observation targeted at the inner Galactic bulge ($l_{\textsc{ii}}$ $\sim$ 0.113$\deg$, $b_{\textsc{ii}}$ $\sim$ -1.424$\deg$), \citet{revnivtsev11} estimate the 2--10 keV X-ray emissivity to be $(2.4\pm0.4) \times 10^{27}$ erg s$^{-1}~\Msun^{-1}$. Finally from a similar \textit{Chandra} study based on deep-field data taken within 1\arcmin-4\arcmin of Sgr A*, \citet{revnivtsev07} derived an X-ray emissivity value of  $\sim (8.5\pm4.3) \times 10^{27}$ erg s$^{-1}~\Msun^{-1}$ (corrected to a GC distance of 8 kpc). Here the large error stems from uncertainty in the GC mass model - see the discussion below.

Taken as a whole, the published measurements provide evidence of an up-lift in the X-ray emissivity per unit stellar mass of the unresolved source population within $< 1$ deg of Sgr A*, which is substantiated by our current results.

Two interpretations are possible for the upward trend in the X-ray emissivity per unit stellar mass within the central 100-pc zone evident in  Fig. \ref{fig:diff}. The first is that some characteristic of the underlying source population is changing in tandem with the stellar density. If we assume the ``base-level'' X-ray emissivity of the old stellar population is roughly $3.0 \times 10^{27}$ erg s$^{-1}~\Msun^{-1}$ ({\it i.e.}, a value representative of the GRXE excluding the GC) and apply this scaling to the projected mass distribution (Fig. \ref{fig:diff}, top-left panel), then the observed X-ray luminosity shows a factor $\sim 2$ enhancement relative to this base-level. Conceivably the binarity of stellar population, and hence the incidence of magnetic CVs, might be dependent on the stellar density. Alternatively, the contribution from other source types ({\it e.g.,} extreme RSCVn binaries) might rise within 50 pc of Sgr A* in a fashion analogous to that inferred for the central stellar cusp \citep{sazonov12}. The nucleus of M31 presents an interesting comparator. Within 1\arcmin~ of the centre of M31, the incidence of luminous X-ray sources ($L_{\textrm{x}} < 10^{36}$ erg s$^{-1}$) increases more rapidly than might be predicted from the K-band light \citep{voss07a}, a result which has been interpreted in terms of the dynamical formation of LMXBs within a dense stellar environment \citep{voss07b}. At the distance of M31, 1\arcmin~ corresponds to a linear extent of 200 pc, a scale-size not dissimilar to the extent of the NSC and NSD components
discussed above. Plausibly, dynamical processes may have given rise to a higher incidence of X-ray emitting CVs and active binaries within the central 100-pc region, than that pertaining in the Galaxy as a whole.

An alternative explanation of the observed upward trend in the X-ray emissivity per unit stellar mass and the implied enhanced X-ray luminosity of the central 100-pc region, is that the mass model we have employed is inaccurate. In fact a number of authors have noted that the uncertainties in the current mass model, including the relative scaling of the NSC and NSD components, may be as high as a factor of $\sim 2$ (\citealt{launhardt02};~\citealt{muno06};~\citealt{revnivtsev07}).

We have investigated what changes in the mass distribution are implied by our measurements, if we make the assumption that the X-ray emissivity within the GC region is constant. Bearing in mind that we are dealing with quantities measured in projection on the plane of the sky ({\it i.e.,} integrated along the line-of-sight through the whole nuclear region), we have necessarily sought a  representative solution rather than a unique one. In that context, we find that the upward trend in the emissivity can be counteracted by steepening, the power-law function describing the decrease with radius of the NSD mass density within $R<120$ pc; specifically the power-law index must be changed from $n_{2}=0.1$ to $n_{2}=1.2$.  The total GC mass profile which results in a roughly constant X-ray emissivity of $3.0 \times 10^{27}$ erg s$^{-1}~\Msun^{-1}$ between 2\arcmin--20\arcmin, is shown in the lower-right panel of Fig. \ref{fig:diff}. We note that compared to the current mass model (\citealt{launhardt02};~\citealt{muno06}), additional mass (roughly a factor 2) is required within 10-100 pc of Sgr A*; however, it is unclear whether this is appropriately assigned to the NSC, the NSD or to a combination of both. In summary, this analysis demonstrates that X-ray tracers may eventually provide a very  effective way of constraining the stellar mass distribution in the GC; of course, this is subject to the caveat that we need to better understand the nature of GC X-ray source population and the evolutionary influences to which it has been subject.

Finally we may use the estimates of this section to illustrate just how extreme the environment is at the centre of the Galaxy.  If we take the stellar mass density to be $\sim 10^{3}$ $\Msun$ pc$^{-3}$ (the value at $\sim 20$ pc on the plane in our revised mass model), then the base-level X-ray emissivity per unit stellar mass assumed above implies an X-ray volume emissivity of $3 \times 10^{30}$ erg s$^{-1}$ pc$^{-3}$. If a typical magnetic CV in this region has an
X-ray luminosity of L$_{\rm{X}}$ $\sim$ 10$^{30}$ erg s$^{-1}$, consistent with the faint end of the source distribution probed in \textit{Chandra} deep-field observations \citep{revnivtsev09}, then the volume density of magnetic CVs is $\sim 3$ pc$^{-3}$. However, only $\sim 5$ per cent   of   all  CVs   are   thought   to   be  IPs   (\citealt{ruiter06}), so the projected number density of all types of CVs rises to 60 pc$^{-3}$, compared to the inferred local density of just  
$\sim 10^{-5} \rm~ $ pc$^{-3}$ (\citealt{warner95}; \citealt{schwope02}).

\subsection{Absence of very-hot diffuse thermal plasma in the GC}

The GRXE can be traced along the Galactic Plane from 
$|l_{\textsc{ii}}| \sim 90\deg$ through to the GC, where its
surface brightness reaches a maximum 
(\citealt{koyama86}, \citeyear{koyama89}; \citealt{revnivtsev06}).
In recent years the debate as to whether this feature is due predominantly
to the presence of a truly diffuse very-hot ($kT \sim 7.5$ keV) thermal plasma
or due to the integrated emission of point sources, has
inexorably swung in favour of the latter interpretation. 

A key recent result in the context of the origin of the emission from the GC
was the discovery that in a deep {\it Chandra} observation
at ($l_{\textsc{ii}}, b_{\textsc{ii}} = +0.08\degn, -1.42\degn$), 
more than $\sim80$ per cent 
of the continuum and line emission around 6.7 keV could be
resolved into point sources \citep{revnivtsev09}. Also detailed spectral
analysis of \textit{Suzaku} observations in the Galactic bulge region, 
suggests that virtually all of the unresolved X-ray emission can be
explained in terms of the integrated emission of point sources \citep{yuasa12}.

In this paper, we have investigated the distribution of X-ray
emission in the region between 2\arcmin~ and 20\arcmin~ from Sgr A*
in a variety of bands, encompassing three iron-line tracers
and also the very-hard 7.2--10 keV continuum.  We find that
in many regions the emission in these bands follows a smooth, symmetric
centrally-concentrated spatial distribution.  Furthermore, to within a factor of 2, the observed surface brightness distribution matches
the projected mass distribution  in the old stellar population
in the GC, specifically that in the Nuclear Stellar Disc
component (\citealt{launhardt02}; \citealt{muno06}). 
We interpret this as strong support for
the conjecture that the observed emission is predominantly
the integrated emission of unresolved point sources.

Excluding the central 2\arcmin, where the Sgr A East SNR, the Central Cluster,
and other features combine to produce a complex morphology
(\citealt{baganoff03}; \citealt{muno04b}), and also a few regions dominated
by individual X-ray binaries or other bright sources
({\it e.g.,} the Arches Cluster), we identify two regions within 20\arcmin~
of the GC where the underlying glow of point sources is 
confused by additional emission components. The first is the
north-east region where X-ray fluorescence and reflection
from dense molecular clouds gives rise to a strong signal in the
6.4-keV iron-line and the very-hard continuum bands.  Excess
emission at 6.7-keV can be seen in this north-east region
and, also in a second ``anomalous'' region to the south of Sgr A*.
In both cases we interpret the 6.7-keV excess
as a concentration of soft thermal emission with temperature
$\sim$ 1.5 keV.  The key point is that within this overall setting
there is no requirement for any excess {\it very-hard} thermal emission 
which might be attributed to a residual diffuse GRXE component. 

In a recent paper, \citet{uchiyama11} used {\it Suzaku} measurements to
trace the distribution of iron-K emission from the
GC proper through to regions of the inner Galactic Plane
($|l_{\textsc{ii}}| = 1\deg-10\deg$).  Similar to the current analysis,
these authors compared the spatial variation of the
Fe \textsc{xxv} K$\alpha$ line with a stellar mass distribution model.
They found that when the X-ray emissivity per unit stellar mass
was fixed at the value required to explain the observed
emission outside of $|l_{\textsc{ii}}| \sim 1.5\deg$, then the
X-ray emission observed within 1\deg was 3.8 times under-predicted.
They concluded that either a new source population with extremely strong
Fe \textsc{xxv} K$\alpha$ emission was required in the GC
or that the majority of hard GC X-ray emission must originate in a
truly diffuse optically-thin thermal plasma. In the present paper we argue that the X-ray emissivity in the GC is roughly a factor 2 higher than estimates pertaining to the local Galaxy and Galactic Plane (\citealt{sazonov06}; \citealt{revnivtsev12}). A possible explanation of these different estimates of the degree to which the GC X-ray emissivity is enhanced  (3.8x as opposed to 2x) is that \citet{uchiyama11} use a scaling for their Galactic disc component, taken from \citet{muno06}, which is based on a total mass in the Galactic disc of $10^{11}~\Msun$. However, as noted by \citet{hong09} this is likely to be an overestimate of the total {\it stellar mass} in the disc; for example \citet{robin03} quote a value of $2.2 \times 10^{10}~\Msun$. Whatever the actual degree of enhancement of the X-ray emissivity in the central 100-pc region, we hold to the view that the observed spatial distribution and spectral properties of this component are best explained in terms of the integrated emission of unresolved sources.

\section{Conclusions}

We have used the extensive set of observations available in the
{\it XMM-Newton} archive to make mosaiced images of the central
100 pc $\times$ 100 pc region of our Galaxy in a variety of
broad and narrow bandpasses within the 2--10 keV energy range.

Using three iron-K lines (Fe \textsc{i} K$\alpha$ at 6.4 keV,
Fe \textsc{xxv} K$\alpha$ at 6.7 keV and Fe \textsc{xxvi} Ly$\alpha$ at
6.9 keV) as tracers, we decompose the GC X-ray  emission into three 
distinct  components.  The first of these components is the integrated
emission of unresolved point sources with 2--10 keV X-ray luminosity $L_{X} <
10^{33} \rm~erg~s^{-1}$.  The second
is the fluorescent line emission and the reflected continuum 
from dense molecular material.  The third is the soft diffuse
emission from thermal plasma in the temperature range,
$kT \approx$ 0.8--1.5 keV.  We show that 
the spatial distribution and spectral properties of the GC X-ray
emission (excluding the central 2\arcmin~ region and a few
other regions confused by luminous sources) can,
to a good approximation, be described in terms of these
three components.

A key finding is that unresolved sources account for much of the
observed emission in 6--10 keV band, including the iron-line
emission.  We find that the surface brightness of this component
falls off as  $\sim \theta^{-0.6}$, where $\theta$ is the
angular offset from Sgr A*, coupled with a latitudinal scale height
factor  $\phi_{sc}$=18.6$^{+1.6}_{-1.2}$ arcmin. 

We assume that the unresolved sources are associated with the
old stellar population of the Galaxy, which can be traced in the
GC through the NIR light, on the basis of which, a 3-d mass model
has been derived by \citet{launhardt02}. According to
\citet{launhardt02} there are two major stellar structures
within the central 100 pc of the Galaxy, namely
the Nuclear Stellar Cluster (NSC) and the Nuclear Stellar Disc (NSD). 
When we project these two components of the mass-model 
onto the 2-d plane of sky, we find that
it is the NSD which dominates the region sampled
by the X-ray measurements.
The X-ray surface brightness
distribution (as parameterised above) rises more rapidly with decreasing
$\theta$ than the
mass-model prediction. One interpretation is  that the 2--10 keV
X-ray emissivity increases from 
$\approx 5 \times 10^{27}$ erg s$^{-1}\Msun^{-1}$
at $20\arcmin$ up to almost twice this value at $2\arcmin$.
Alternatively, some refinement of the mass model may be required. Specifically if we assume the X-ray emissivity in the central region is similar to that pertaining to GRXE as a whole (\textit{i.e.}, $\sim 3 \times 10^{27}$ erg s$^{-1}\Msun^{-1}$), then roughly a factor of two additional mass is required within 10--100 pc of Sgr A*. However, it is unclear whether this applies to the NSC, the NSD, or a combination of both components.

 When we set $n_{2}=1.2$ in the NSD model
and revise the model normalisation 
so as to conserve the total mass in the NSD within our
2\arcmin--20\arcmin~field of view, we obtain an X-ray emissitivity
per unit stellar mass of $\approx 6.5 \times 10^{27}$ erg s$^{-1}\Msun^{-1}$. 
This is within a factor 2 of 
other estimates of this quantity based on the local X-ray luminosity
function and studies of the GRXE component outside of the GC region.

The unresolved hard X-ray emitting source population,  on the basis
of  spectral  comparisons,  is   most  likely  dominated  by  magnetic
cataclysmic  variables, primarily intermediate polars.
By way of example,  we  use  the  X-ray   observations 
to estimate that there are 60 CVs (of all types) per cubic parsec
at a radial distance of 20 pc from Sgr A*. This is close to seven orders of magnitude over the local CV number density.

A further major finding of our work is that our composite
emission model, encompassing the three components described above,
provides a sufficiently good description of the observations
so as to rule out any substantial additional source of
hard X-ray luminosity in the GC. It has long been conjectured
that a  significant fraction of the hard X-ray emission from the GC
originates in very-hot ($\sim7.5$ keV) {\it diffuse} thermal plasma.
Within our modelling framework, the only way of accommodating a
substantial contribution of this nature,
is within the emission component we attribute to
unresolved sources.  However, the properties of the latter, namely
the smooth, symmetrical, centrally-concentrated spatial distribution,
an X-ray emissivity reasonably in accord with local estimates,
and an X-ray spectrum matching that of intermediate polars,
would seem to exclude the diffuse, very-hot thermal plasma
hypothesis as a viable option.

\section*{Acknowledgments}

VH  acknowledges  the financial  support
provided by the UK STFC research council.
This work is based on \textit{XMM-Newton} observations, an ESA mission
with  instruments  and contributions  directly  funded  by ESA  member
states  and the  USA  (NASA). 

\bibliography{diffuse_emission_refs_rev1}

\bibliographystyle{mn2e}

\bsp

\label{lastpage}

\end{document}